\documentclass{emulateapj}

\slugcomment{}%

\begin{document}

\shorttitle{Near-IR CO Emission in the Hypergiant $\rho$ Cas}
\shortauthors{Gorlova et al.}


\title{On the CO Near-IR Band and the Line Splitting Phenomenon in the Yellow Hypergiant $\rho$ Cassiopeiae}


\author{Nadya Gorlova$^1$, Alex Lobel$^2$, Adam J. Burgasser$^3$, George H. Rieke$^4$, Ilya Ilyin$^5$, and John R. Stauffer$^6$ }

\affil{$^1$Steward Observatory, University of Arizona, 933 North Cherry Avenue, Room N204, Tucson, AZ 85721-0065} \email{ngorlova@as.arizona.edu}
\affil{$^2$Royal Observatory of Belgium, Ringlaan 3, 1180 Brussels, Belgium} \email{alobel@sdf.lonestar.org}
\affil{$^3$Massachusetts Institute of Technology, Kavli Institute for Astrophysics and Space Research, 77 Massachusetts Ave, Building 37, Cambridge, MA 02139}\email{ajb@space.mit.edu}
\affil{$^4$Steward Observatory, University of Arizona, 933 North Cherry Avenue, Room N204, Tucson, AZ 85721-0065} \email{grieke@as.arizona.edu}
\affil{$^5$Astrophysical Institute Potsdam, An der Sternwarte 16, 14482 Potsdam, Germany} \email{ilyin@aip.de}
\affil{$^6$Spitzer Science Center, California Institute of Technology, MS 314-6, Pasadena, CA 91125}\email{stauffer@ipac.caltech.edu}



\begin{abstract}
We report on multi-epoch optical and near-infrared spectroscopy around the first overtone 
ro-vibrational band of CO in the pulsating yellow hypergiant $\rho$ Cas, 
one of the most massive stars in the Galaxy and a candidate SN II progenitor. 
We argue that the double cores of the CO absorption lines, that have previously been 
attributed to separate circumstellar shells expelled during its recurrent outbursts, 
result in fact from a superposition of a wide absorption line and a narrow central 
emission line. The CO line doubling returns over subsequent pulsation cycles,
where the superposed line emission assumes its largest intensity near phases of maximum light.
We find that the morphology and behavior of the CO band closely 
resemble the remarkable "line-splitting phenomenon" also observed in optical 
low-excitation atomic lines.
Based on radiative transport calculations we present a simplified model of the near-IR 
CO emission emerging from cooler atmospheric layers in the immediate vicinity of the photosphere. 
We speculate that the kinetic temperature minimum in our model results from a 
periodical pulsation-driven shock wave. We further discuss a number of alternative
explanations for the origin of the ubiquitous emission line spectrum, possibly due to a 
quasi-chromosphere or a steady shock wave at the interface of a fast expanding wind 
and the ISM. We present a number of interesting spectroscopic similarities between 
$\rho$ Cas and other types of cool variable supergiants such as the 
RV Tau and R CrB stars. We further propose a possibly common mechanism 
for the enigmatic outburst behavior of these luminous pulsating cool stars.
\end{abstract}






\section{Introduction}

$\rho$ Cas (HD224014) is a naked-eye variable supergiant star 
($V$ = $4^{\rm m}.5$, $K$ = $2^{\rm m}$)
with a long history of photo-visual photometric and spectroscopic observations.
It was observed with the first near-infrared (NIR, 1--3$\mu$m) spectrographs in the
late 1970s \citep{Lambert81}, but has been somewhat neglected since then. 
This paper discusses recent spectroscopic monitoring of the NIR spectrum of this 
remarkable hypergiant star, revealing a wealth of remarkable temporal spectral features 
whose interpretation remains elusive. For example the cores of numerous strong 
absorption lines always appear double. The apparent line core splitting results from a 
static central emission component. However in certain variability phases 
$\rho$ Cas also reveals a prominent optical and NIR emission line spectrum that can indicate 
a common physical origin with the central emission components of split absorption lines.    
An important question to further address the peculiar nature of $\rho$ Cas' emission 
spectrum is therefore if the central emission lines and the prominent emission
line spectrum form in similar circumstellar conditions (e.g. at similar distances from 
the surface of the hypergiant with comparable geometric density distributions).
Detailed spectroscopic studies of the circumstellar environments 
of cool hypergiant stars are scarce, although they are absolutely required to properly 
understand the physical mechanisms that can explain their uncommonly large observed mass-loss 
rates. For example, can the large mass-loss rates observed in F- and G-type hypergiants, such as 
$\rho$ Cas (that are too cool to accelerate the radiatively line-driven winds of luminous hot 
stars), result from atmospheric pulsations, since their extended winds are too warm to be
accelerated by dust-driven mechanisms commonly proposed in red supergiants?
Can the observed properties of $\rho$ Cas' variable emission line spectrum be 
reconciled with the pulsation kinematics of the extended atmosphere and with its uncommonly 
large mass-loss rates?

With a spectral type of F8-G2 Ia0 ($T_{\rm eff}$ = 6500-7200 K), and a 
luminosity of log $L_{*}$/$L_{\odot}$ $\sim$ 5.7 ($R_{\star}$ $\simeq$ 400-500 R$_{\odot}$), 
$\rho$ Cas is one of a small sample of known yellow hypergiants in 
the Galaxy. In quiescent variability phases the star looses mass
at a rate of few $10^{-4}M_{\odot}$, which is intermediate for cool hypergiant stars, but 
can increase by two orders of magnitude during rare outburst episodes \citep{Lobel98,Lobel2003}.
Evolutionary tracks (e.g. \citet{Lejeune2001}) and possible membership
to the Cas OB5 association \citep{Humphreys78} indicate a stellar mass of 
$\sim$40~M$_{\odot}$ and possible ages of 4--6 Myr.
Observations of CNO-processed abundances in its extended 
Na-enriched atmosphere \citep{Takeda94} signal that this yellow supergiant has gone 
through dredge-up in the red supergiant stage, and is presently evolving bluewards. 
The star is about to cross the so-called atmospheric instability region in the upper 
HR-diagram: "the Yellow Evolutionary Void" \citep{Nieuwenhuijzen1995}, where due to 
the enormous atmospheric scale height of these massive evolved stars, coupled with 
a large temperature in the shell-burning layers, the effective surface acceleration 
essentially vanishes, producing dynamically unstable stellar atmospheres. 

$\rho$ Cas's atmospheric instability is manifested in both its quasi-periodic 
photometric behavior and unusual spectral variability. It normally pulsates 
semi-regularly (Srd) with $P$ $=$ 320-500 d, and an amplitude of 
$\pm$ $0^{\rm m}.2$ ($V$ band) \citep{Zsoldos91,Arrelano1985}. 
Once every 20 to 50 years, however, this hypergiant goes into an outburst by 
dimming more than a full visual magnitude, decreasing the $T_{\rm eff}$ by 
more than 3000 K, and by ejecting a massive gas shell. The most recent 
outburst event occurred in 2000-01 which was carefully monitored using 
high-resolution optical spectroscopy \citep{Lobel2003}. In this paper we 
present the NIR (2 $\mu$m) spectrum from that exceptional epoch in \S \ref{atem}.

The spectrum of $\rho$ Cas outside outburst is peculiar. Optical studies 
reveal that the spectral lines are very broad due to large atmospheric 
micro- and macro-turbulence velocities (11 and 21 $\rm km\,s^{-1}$, respectively). 
The detailed profiles of certain lines are also rather complicated.  
Besides the single-component weak absorption ("normal") lines that are entirely 
formed in the supergiant's lower photosphere, the low-excitation ($\chi_{\rm low} \lesssim$ 3 eV) 
absorption lines are either {\it cyclically} or {\it permanently} split 
(depending on the line oscillator strength within a multiplet), having one 
component always blue- and the other one red-shifted relative to the 
stellar radial velocity of $-$47$\pm$1 $\rm km\,s^{-1}$.
A number of low-excitation lines appear in emission above flux level of the 
local continuum around variability phases of maximum light (e.g. when 
$T_{\rm eff}$ becomes largest), while others, like the forbidden [CaII] lines, 
peak always prominently above the local continuum level throughout a complete variability cycle.
In addition, many medium-strong and strong absorption lines develop far violet extended 
wings during certain phases of fast wind expansion. They can sometimes 
also display a rather "bumpy structure" in the far short-wavelength wing 
profile of split absorption lines. 

Despite the fact that $\rho$ Cas's peculiar spectrum was immediately recognized
almost a century ago, its precise physical origin remains heavily debated.
This is no surprise given that it requires an appreciable amount of 
observing time to acquire the substantial body of simultaneous spectroscopic and 
photometric data necessary to reveal a correlated pattern of behavior. The earliest interpretations 
of the peculiar line splitting phenomenon attribute 
it to two detached circumstellar shells previously ejected by the supergiant 
\citep{Bidelman55, Sargent61, Gesicki92}.
Various authors remarked that such shells possibly comprise two atmospheric 
layers flanking an outwardly 
propagating shock wave, a mechanism currently used to explain the absorption line 
splitting observed in a number of less-luminous variables such as RR Lyr, 
W Vir, RV Tau, and Miras \citep[e. g.,][]{Chadid96,Whitney62,Gillet89,Hinkle82}. 
\citet{Sheffer86} added support to the shock 
wave hypothesis for $\rho$ Cas by showing that the line core doubling correlates 
with the pulsation phase, because the splitting is most prominent around maximum 
light when the photosphere begins to expand. They also argued that the line core 
doubling first starts in the high-excitation lines, as if a shock wave were 
propagating outwards crossing various atmospheric regions with decreasing 
$\chi_{\rm low}$. However, \citet{Sheffer93} also considered a different 
explanation for the split lines of $\rho$ Cas. Instead of two separate 
absorption components, the splitting could be interpreted as a superposition of
a single broad absorption line and a static central \textit{emission} line.
This proposition was based on the observation of the cores of some split lines 
that become so intense at times that they peak above the local level of the 
continuum, sometimes without any accompanying absorption, which also rules 
out a P Cygni-type line formation region in $\rho$ Cas. These effects were also 
noticed for the slightly warmer hypergiant HR~8752. The emission can be better
detected against the weaker absorption spectrum of this hypergiant \citep{Sheffer87}. 
\citet{Sheffer87} hence proposed that the origin of the variable emission 
lines could be in a putative extended chromosphere around these very luminous 
cool stars. 

Further steps at discarding the "shell hypothesis" for the formation of 
split absorption lines in $\rho$ Cas were provided in \citet{Lobel98}. 
They showed that 1) the observation of {\it permanently} split absorption lines in 
the optical spectrum, with components remaining at a constant radial velocity, would 
require two absorbing 'shells' moving with positive and negative radial velocity 
during the complete variability cycle. Two separate shells permanently propagating 
in opposite directions is physically excluded;
2) the high-excitation lines considered in \citet{Sheffer86}
do not necessarily form deeper than low-excitation lines in the extended atmosphere 
of $\rho$ Cas, and therefore do not provide additional 'evidence'
for an outwardly propagating atmospheric shock wave;
3) the variations of the red- and blue-shifted components in the split absorption 
lines are not independent of each other. One line component always grows stronger
while the other line component weakens. This effect can readily be mimicked by 
assuming a central static emission line superimposed on a stronger and broader 
absorption line that exhibits small Doppler displacements due to the photospheric 
oscillations. Further direct evidence against substantial circumstellar material 
around $\rho$ Cas includes the non-detection by HST-WFPC2 of any significant 
structure at 0.1\arcsec\, or 310 AU $\approx$ 150 $R_{\star}$ away from the 
hypergiant \citep{Schuster05}. Nevertheless, tenuous and temporal circumstellar 
material may still be present as indicated by a weak mid-IR excess \citep{Jura90}
and the spectroscopically observed mass ejections during recurrent 
outbursts of the star (e.g. optical TiO band observations by \citet{Lobel2003} 
and references therein to the two earlier recorded outbursts).

In this paper we address the important question of the origin and nature 
of the line emission region in $\rho$ Cas. In particular, what are its thermal 
conditions and what is the geometrical extension and location of this region? 
Can the emission region and its line excitation mechanism be linked to the 
pulsations of the lower photosphere of the hypergiant? We discuss several 
important clues from NIR spectra of $\rho$ Cas, but will also scrutinize 
new high-resolution optical and UV observations
(of which a more detailed analysis will be presented elsewhere). 
 
While optical TiO bands have only been recorded during rare outbursts,
the abundant CO molecule is however always observed in the NIR spectrum of $\rho$ Cas. 
The NIR spectrum was extensively monitored by Sheffer et al. (1993) 
with the FTS spectrograph on the Mayall 4m telescope (Kitt Peak) from 1979 
to 1989. Their intense monitoring program revealed substantial 
NIR spectral changes on a time scale of only months \citep{Sheffer93}.
Although the "shell hypothesis" can be abandoned for the optical split 
absorption lines, these authors adopted it to interpret the CO line variability 
in the NIR spectrum. Unlike the optical atomic split lines, the NIR CO lines 
require kinetic gas temperatures much smaller than those adopted in classical 
photospheric models for F- and G-type supergiants
(for a comprehensive CO Grotrian diagram see Fig. 1 of \citet{Bieging02}).
The low kinetic gas temperatures ($\lesssim$ 2000 K) required 
for CO molecular formation at these 
early spectral types either require a temperature reversal in the upper 
atmosphere (a model we explore in \S \ref{shock}), or must assume the presence 
of a much cooler circumstellar gas envelope. The latter interpretation was 
adopted by \citet{Sheffer93} by attributing the short-wavelength "shell" 
component of the split CO absorption lines to a moderate outburst of $\rho$ Cas
in 1986. They further assumed the detached shell interpretation to compute 
the total mass of the expelled gas layer, its turbulent velocity, 
and the atmospheric deceleration during their monitoring period. 

NIR CO bands have been observed and interpreted as direct tracers 
of ejected stellar material in a number of other F-G supergiants.
For example, \citet{Oudmaijer95} discuss the molecular winds of post-AGB stars, 
while \citet{Gonzalez98} investigate dust formation close to the stellar 
surface of R CrB stars. Given the difficulties to properly explain the 
properties of split {\it atomic} lines in $\rho$ Cas, we argue that 
a ``shell hypothesis'' for CO line formation needs to be re-examined as well.
The early investigations by Sheffer already indicated that despite the wide 
variety of morphologies exhibited by the CO line profiles, they can nonetheless 
be reduced to a narrow central emission  line superimposed on a broad 
absorption line, yielding two adjacent (but therefore only "apparent") 
absorption line components. However, their study was not widely published 
or discussed, and only one FTS spectrum has been discussed by \citet{Lambert81}. 
The major part of these exceptional NIR spectroscopic data were only 
presented as part of a Ph.D. thesis \citep{Sheffer93}.

We present a collection of NIR spectra of $\rho$ Cas obtained with several
telescopes in 2000-04. Our NIR monitoring program is not as frequent as Sheffer's, 
but has the following benefits. First, we present spectral analyses of 
new NIR observations in combination with our ongoing optical high-resolution 
spectroscopic monitoring of $\rho$ Cas since 1993 \citep{Lobel2003}.
Next, while Sheffer et al. studied NIR spectra 
of the moderate outburst of 1986, we investigate the major 
outburst event of 2000-01. Following the latter major outburst
the long-term variations in H$\alpha$ reveal the formation of an exceptional 
P Cygni type line profile since the spring of 2004 \citep{Lobel2004, Lobel2005}. 
Cool supergiant stars often exhibit secular variations on time scales
of decades (for examples see \S \ref{hyps}) and it is therefore important
to continue accumulating spectroscopic observations of these 
unusual stars.

Although our combined spectra span the entire range from
optical wavelengths longward of H$\alpha$ through the main NIR 
$J$, $H$, and $K$ bands, up to $L$ (4 $\mu$m), the present study 
primarily concentrates on the $K$ band region where the first 
overtone band of CO occurs (\S \ref{covar}). Our high-resolution 
CGS4 spectral observations of 2004 September are motivated by 
the large variability of this molecular band observed in low-resolution spectra,
and because it was detected in strong emission in 1998.
We also draw attention to the $L$ band region in the 
SpeX spectrum (\S \ref{atem}) because, to our knowledge, it is 
the first $L$ band spectrum of medium spectral resolution 
presented for any F-type supergiant like $\rho$ Cas.

The outline of this paper is as follows. First we describe
new optical to NIR spectra of $\rho$ Cas 
obtained in 1998-2004, covering roughly four pulsation cycles (\S \ref{obs}).
Then we establish the presence of emission in NIR CO lines and investigate 
their profile changes with the stellar pulsations (\S \ref{covar}).
We discuss a unified excitation mechanism that can produce emission in 
both molecular and atomic lines, considering several 
explanations based on detailed CO line modeling, combined 
with co-eval optical spectroscopy (\S \ref{comod}). Finally,
we compare our observations with 
the line splitting phenomenon reported in other cool star spectra. 
For example, we show that brightness declines 
of R CrB stars appear intriguingly similar to the outbursts of $\rho$ Cas (\S \ref{discus}). 
The comprehensive approach of our NIR monitoring program enables us to 
investigate the remarkable evolution of the CO 2.3 $\mu$m band, and to provide 
credible evidence for the stellar pulsations as the dominant factor that determines 
the profile changes observed in NIR and optical low-excitation spectral lines (\S \ref{concl}).

\section{Observations and Data Reduction}\label{obs}

\subsection{IR Spectroscopy}

We have collected a series of low ($R$ $\sim$ 300)
to high ($R$ $\sim$ 40,000) spectral resolution NIR spectra 
of $\rho$ Cas covering roughly four pulsation cycles in 1998-2004.
Our observations utilize three instruments on three telescopes --
the CorMASS echelle spectrograph on the 1.9m Vatican Advanced 
Technology Telescope in Arizona \citep{Wilson01}, the SpeX echelle 
spectrograph on the 3m InfraRed Telescope Facility \citep{Rayner03}, 
and the CGS4 spectrograph on the 3.8m United Kingdom Infra-Red Telescope 
\citep{Mountain90} at Hawaii.
Observations with the latter spectrograph were obtained in 
service mode. The datasets are complemented with two archival 
CGS4 spectra, one obtained in 1998, one pulsation cycle before,
and the other one right in the middle of the grand outburst minimum of 2000-01. 
Table \ref{table1} summarizes all observing dates and the 
basic characteristics of the new NIR spectra.

All NIR observations are performed in a standard fashion resulting in 
similar data reduction procedures. A series of successive A-B sequences 
(where A and B correspond to two positions of the star along the slit) 
are obtained, both for $\rho$ Cas and a bright telluric standard A-, F-, 
or G-type star. Exposure times range from 0.5 to 2 s, usually restricted 
to the shortest time allowed for a particular instrument. 
The size of the entrance slit ranges from 0.3\arcsec\, to 2\arcsec, the
allowed minimum to attain maximum available spectral resolution. 
To suppress the brightness of $\rho$ Cas before saturation a neutral 
density filter is used with CorMASS and for the low-resolution mode of CGS4 
(both for $\rho$ Cas and telluric standards).

The flat-fielded A and B images are combined and mutually subtracted 
to perform dark and sky removal. The spectral orders are extracted, 
wavelength-calibrated, and divided by a standard star spectrum 
with removed intrinsic hydrogen lines. Next they are trimmed 
and combined into a single NIR spectrum.
The resulting spectrum is then
multiplied by a black-body spectrum corresponding to the spectral type 
of the standard. The wavelength calibration of the 
CorMASS spectra is performed using observations of the planetary nebula 
NGC 7027 obtained during the same night or the night before. 
The calibration of the high-resolution CGS4 spectra and the 
short-wavelength mode of SpeX employ Ne-Ar-Xe arc lamp exposures 
obtained immediately after science target observation. The low-resolution 
CGS4 observations and the SpeX long-wavelength mode calibration both employ 
these arc lamp exposures as well as extraction of sky lines from the 
target images. An accurate absolute flux calibration
has not been attempted for the obvious reasons of intentional
flux suppression and strong seeing-induced variations between 
subsequent exposures of this very bright star.

A number of important differences in our reduction procedures 
using different instrument/grating settings are the following. 
The 40 l/mm CGS4 observations require one additional step of correcting
for a 2 to 3 pixel-wide wiggle in the flux level introduced by 
variations in the seeing conditions. It occurs when the detector 
is shifted in the dispersion direction between subsequent exposures
to improve the sampling of the resolution element on the detector array.
The basic steps for reduction of the calibration frames and co-addition 
of the target frames are performed with the {\tt ORAC-DR} pipeline for 
the CGS4 low-resolution spectra. The {\tt SpeXtool} software package is used
for the SpeX data which also performs the extraction of the final 
1-d spectra \citep{Vacca03, Cushing04}. The CorMASS spectra are 
reduced with the standard {\sc iraf} software package {\tt imred.echelle}.
The {\tt SpeXtool} routine {\tt Xtellcor} provides a modified spectrum of Vega
to perform telluric corrections in the SpeX spectra. For all other cases 
strong hydrogen lines in the standard spectra are interpolated
with a straight line and removed using the {\sc iraf} task {\tt splot}.

\subsection{Optical Spectroscopy}

In order to investigate correlations between behavior of CO lines,
the photosphere, and the circumstellar components observed in atomic lines, we complement
Table \ref{table1} with seven
high-resolution optical echelle spectra obtained during our continuous 
monitoring program of the star in 2000-04 (\S \ref{atomic}). 

The spectrum of  2000 July 19 was observed during the deep outburst minimum with 
the Utrecht Echelle Spectrograph on the 4.2m William Herschel Telescope 
at the Canary Islands (Spain). The nominal wavelength resolution is
$R$ $\simeq$50,000, with continuum signal-to-noise ratios exceeding 100. 
For more detailed discussions 
about calibration procedures and a thorough analysis of the outburst 
spectra we refer to \citet{Lobel2003}.

The spectrum of 2004 October 23 was observed to follow up our
CGS4 NIR observations of September 2004 that revealed strong 
CO emission. It was obtained 5 weeks later, and about a week before the 
CorMASS NIR observations, when the remarkable CO emission had already started 
to decrease. The spectrum was obtained with the KPNO 4m Mayall
telescope using the Echelle Spectrograph configured to provide
$R$ $\approx$40,000 in the red (6000-8500 \AA\, in 22 orders).
The wavelength calibration was performed with the Th-Ar lamp spectra,
with a small additional shift in the zero-point
as inferred from terrestrial $\rm O_{2}$ lines.

The other optical echelle spectra of 2003-04 were observed with the Sofin 
spectrograph on the 2.6m Nordic Optical Telescope (La Palma, Spain).
The spectra are observed with the medium-resolution Camera 2  
($R$ $\simeq$ 80,000) with a wavelength range covering about 5000 \AA\,
over 42 echelle orders. Each observation combines a number of subsequent 
exposures with two central wavelength settings until a minium S/N ratio of $\sim$50 
is reached. The S/N ratios for the echelle orders centered on the CCD can 
therefore exceed 300 in the wavelength region of the spectral lines of 
interest. Typical effective exposure times are limited to 10 min. per 
exposure to avoid CCD saturation for this optically bright star. 
Note for example that the large brightness of $\rho$ Cas allows us to 
integrate high-quality optical spectra during dusk or dawn (i.e. at the beginning
or end of a bright night run) with large zenith distances ($\geq$50 $\deg$).
For some of our spectra it yields large air-masses (e.g. $\sim$1.6 
for 2004 December 25) with strongly contaminating telluric lines, but which 
otherwise do not influence our spectral regions of interest.      
The Sofin spectra are calibrated with Th-Ar lamp spectra observed 
after each science exposure. The standard echelle reduction steps are 
performed using the 4A data acquisition and reduction software package
described in \citet{Ilyin2000}. Special consideration is given to the removal
of scattered light inside the spectrograph, and to the extraction of 
spectral orders including the removal of cosmic spikes. The wavelength
dispersion solutions yield mean accuracies better than $\pm$1~$\rm km\,s^{-1}$.
The echelle spectra are not calibrated in an absolute flux scale, but 
special care is taken to apply the required blaze corrections for an 
accurate placement of continuum flux levels for continuum normalization. 
The local continuum levels are either determined with a polynomial fit 
through pre-determined anchor points in wavelength regions that are relatively 
free of spectral lines, or are based on a best visual estimate systematically applied 
to all of our observations. 
Note that for these large S/N ratio optical 
spectra the accuracy of the placement of stellar continuum level
is better than 1-2\%, which is sufficient for measuring the small intensity 
changes we observe over time in weak absorption and emission lines (\S \ref{atomic}).

\section{Remarkable Variations of $\rho$ Cas Near-IR Spectrum}\label{covar}

\subsection{Spectra near Minimum and Maximum Light}\label{atem}

Due to a persistent lack of systematic libraries in the literature
of NIR spectra of warm supergiant stars, we observed CorMASS spectra of
several bright supergiants for comparison with $\rho$ Cas.  
The spectra were obtained in 2003-04 with a setup 
identical to that for $\rho$ Cas.
Figure \ref{fig1} shows our low-resolution NIR spectra of $\rho$ Cas with 
less-luminous supergiants of similar spectral type: the non-variable 
$\mu$ Per (G0 Ib) and the 3d-period classical Cepheid Y Aur (F-G) 
in two variability phases. HR 8752 is another semi-regular hypergiant
which was the spectroscopic twin of $\rho$ Cas in the 1970s but whose  
$T_{\rm eff}$ increased to above 8000 K since then \citep{Israelian99}.
Due to the low resolution of CorMass
it is difficult to determine if HR 8752 currently is an early A or F star,
although it clearly reveals a phase with strong emission observed in
H$\alpha$, Pa$\beta$ and the Ca~{\sc ii} IR triplet lines.
The latter confirms the strong emission we observe 
in the violet wings of the IR Ca~{\sc ii} lines in 
1998 October with NOT-Sofin, although these lines also reveal a central strong 
absorption component in the higher resolution spectra.
The H$\alpha$ line in HR 8752 has a 
permanent triple-peaked emission profile which, unlike in $\rho$ Cas, may 
be attributed to the presence of a faint B-type companion discovered with $IUE$
in HR 8752. On the other hand however, the broad Ca~{\sc ii} H \& K resonance 
lines of both hypergiants lack strong central emission cores, signaling 
the absence of a classical chromosphere in these cool hypergiants.
The more detailed comparison between the two hypergiants is given in \S \ref{hyps}.
 
The overall spectral energy distribution of $\rho$ Cas beyond 1 $\mu$m 
compares well with what is expected for a F-G -type star (basically 
a black-body of $T$ $\sim$ 6000-7000 K). It confirms the small 
reddening of only $A_{\rm V}$=$1^{\rm m}.3$ (\citet{Zsoldos91}), of interstellar 
origin. The reddening is barely noticeable with a slight flattening of the 
spectrum shortward of 0.8 $\mu$m. There is no substantial 
NIR excess around 2 $\mu$m as could be expected from warm circumstellar 
dust near the sublimation temperature of $\sim$1500 K.
The low surface gravities (log $g$) of $\rho$ Cas and HR 8752 are 
evidenced by strong O~{\sc i} 0.777 $\mu$m triplet lines in the low-resolution 
spectra. Other enhanced lines observed in $\rho$ Cas around 1 $\mu$m
belong to high-excitation transitions of C~{\sc i} and S~{\sc i}, 
and to low-excitation transitions of Sr~{\sc ii} (the latter is also 
an atmospheric gravity indicator). These lines can signal high excitation 
conditions due to the low gas densities of hypergiant atmospheres,
as well as a possible metal richness for $\rho$ Cas 
($\Delta$ log $Z$=$0.4\pm0.3$; \citet{Lobel94}). 

The $K$ band spectrum of $\rho$ Cas also shows 
CO absorption. CO NIR bands usually occur only in stars of 
K-type or later \citep{Kleinmann86,Wallace96,Wallace97} and are indeed not 
detected in any other yellow supergiant of our sample.
Even more astonishing is the presence of strong CO band {\it emission}
in 1998 October (CGS4 archival spectrum)
when the optical spectra of \citet{Lobel2003}
reveal atomic emission lines.
The CO emission can hence be considered as the continuation of $\rho$ 
Cas' emission line spectrum from optical to infrared wavelengths,
prompting a more detailed search for other NIR emission lines besides 
CO and Na~{\sc i}.

In 2004 July we observe again a number of emission lines in the optical spectrum.
We also observe the emission lines, including CO, one month later with a 
medium-resolution SpeX observation. The most obvious emission features are
marked in Fig. \ref{fig2}.
We identify them as due to a low excitation energy transition
(below 3 eV) of Ti~{\sc i} at 1.04 $\mu$m, while the lines at 1.5 and 1.7 $\mu$m
of Mg~{\sc i} and of Na~{\sc i} at 2.2 $\mu$m emerge from considerably higher
energy levels above 3 eV in the atom.
NIR emission lines are more easily detected in F-G stars toward longer wavelengths
because the continuum flux is weaker and is devoid of strong absorption features.   
We indeed observe a number of emission lines in the $L$ band. They are marked in Fig. \ref{fig2} 
and wavelengths and fluxes are listed in Table \ref{table2}. These lines cannot be identified
with molecular features observed in 
cool giants, such as SiO, HCl, OH, or HCS \citep{Ridgway84,Wallace02}.
Because of the sharp line profiles we think they are of atomic origin. 
We searched for candidate transitions with atomic line properties comparable to the emission lines 
we already identified in the optical and NIR spectrum. We queried the Atomic Line 
List database compiled by van Hoof (URL: {\tt www.pa.uky.edu/$\sim$peter/atomic})
but obtain for each emission line more than one candidate atomic transition.
One resolution element in our $L$-band spectrum equals 
about four times the average line width of emission lines
(FWHM $\lesssim$ 30 $\rm km\,s^{-1}$ as measured from the optical and NIR 
atomic emission lines).
We therefore refrain from further NIR emission line identifications until a 
higher resolution spectrum becomes available. 

We observe that the NIR emission spectrum appears at least twice simultaneously
with optical emission phases that also coincide with phases of maximum light.
Next we discuss the NIR spectrum for a phase of \textit{minimum} light, which also occurs
during the unusual outburst epochs.
Figure \ref{fig3} shows a NIR UKIRT spectrum of $\rho$ Cas during the outburst in 2000 October.
It is observed in the deep visual brightness minimum when the star was $\sim$1 magnitude 
fainter. To our knowledge it is the first IR spectrum of $\rho$ Cas observed 
during a major outburst event.
Compared to our SpeX spectrum observed around maximum brightness in September 2004,  
emission lines are not observed in outburst except for Na~{\sc i} 2.2 $\mu$m, whereas the 
absorption spectrum is much stronger (except for Br$\gamma$). The Na~{\sc i} emission 
lines protrude through underlying absorption lines. The spectrum 
is overall compatible with the photospheric spectrum of an early K-type supergiant. 
The central depths of CO, Ca~{\sc i}, and Al~{\sc i} lines, and of the Na~{\sc i} 
absorption lines (and possibly also a CN band but which is in the noisy 
region of telluric absorption) compare well with the K4-type supergiant 63 Cyg in Fig. \ref{fig3}.  
This is in agreement with a $T_{\rm eff}$ estimate of $\simeq$4250 K obtained for the optical outburst 
spectrum in \citet{Lobel2003}.
The outburst spectrum reveals however a number of interesting peculiar properties.
Br${\gamma}$ is unusually strong for a K-type star with a central line depth only
slightly weaker than in the SpeX spectrum of $T_{\rm eff}$ $\simeq$7200 K.
On the other hand, H${\alpha}$ is considerably weaker during outburst \citep{Lobel2003}.
The low-excitation ($\chi_{\rm low} < $2 eV) Ti~{\sc i} absorption lines at 2.17-2.23 $\mu$m
are also very remarkable. They appear somewhat stronger than in K- or G-type stars, 
in particular the 2.178 $\mu$m line that compares better to the M2-type supergiant 
$\mu$ Cep with $T_{\rm eff}$ $\simeq$3500 K.
It is of note that these lines have been observed in emission 
in $\rho$ Cas during a shallow brightness minimum of 1980 June \citep{Sheffer93}.
They were also observed in emission for a short period of about one month 
during the eruption of the famous light-echo peculiar supergiant V838 Mon,
while the object assumed minimum brightness and NIR CO was in absorption 
\citep{Banerjee02, Lynch2004}. 
We therefore propose that the strong Ti~{\sc i} lines in the 
NIR UKIRT spectrum of $\rho$ Cas can originate from a 3000-4000 K gas envelope 
ejected during the strong 2000-01 outburst \citep{Lobel2003}.

\subsection{Two-component CO Lines}\label{cotwo}

We investigate the possible correlation of the NIR CO 
variability with the brightness curve utilizing all spectra available to us.
The visual light curve in Fig. \ref{fig4} is 
based on observations by the American Association of Variable
Star Observers (AAVSO). Although their dataset is 
less precise than other photo-electric observations, 
it is the only one available that continuously monitors 
$\rho$ Cas from its earliest spectroscopic observations until 
our most recent observations of 2004. We utilize the visual data
to roughly estimate the variability phase (maximum vs. minimum light, 
brightness increase vs. decrease) as Srd variables are not strictly 
periodic, but only 'quasi-periodic'. Note that the light curve of this 
pulsating star follows the radial movements of its photosphere. 
\citet{Lobel2003} showed that the visual brightness changes mimic the 
radial velocity curve of $\rho$ Cas after 1997, with an offset of about a 
quarter of a period. The atmosphere begins to expand near maximum light 
and reverses to contract after minimum light, for example as is also observed in 
more regular pulsating stars such as classical Cepheids. Unfortunately, 
radial velocity measurements require high-resolution spectroscopic
monitoring which is not available for all epochs of our observations.
We therefore use the visual light curve as a proxy for determining 
the corresponding pulsation phases. 

Figure \ref{fig4} plots the low-resolution $K$ band spectra of $\rho$ Cas against 
the light curve. We find that the CO band always appears in emission near phases of 
maximum light (when the stellar radius already begins to expand) and that the band 
returns to absorption soon after the star dims. The CO absorption in the outburst 
spectrum of 2000 (shown in detail in Fig. \ref{fig3}) is only moderately strong,
which is somewhat surprising considering the deep brightness minimum. 
However, at the low spectral resolution one can not dismiss the possibility of 
filling-in by stronger than usual (residual) CO emission. Alternatively, the 
weakness of the CO absorption can be explained if the non-photospheric
upper layers have been perturbed, if not entirely disrupted by the eruption.

We further concentrate on the recent emission epoch of 2004 with the NIR echelle spectra.
The emission epoch started around 2004 May (based on optical emission spectra 
\citep{Lobel2005}) and ended around October. The resolution of the SpeX spectrum is too 
low to resolve individual line profiles of $\rho$ Cas, although it provides a very 
large wavelength coverage. We therefore obtained a CGS4 echelle spectrum, with 
R$\sim$8 $\rm km\,s^{-1}$ per two pixels, covering five spectral regions centered 
on Br${\gamma}$, Na~{\sc i} 2.2 $\mu$m, and  the first overtone $\Delta \nu$ 2-0 band of CO, 
from the bandhead at 2.29 $\mu$m to 2.322 $\mu$m, where the band begins to overlap with 
the $\Delta \nu$ 3-1 CO band. The SpeX and CGS4 echelle observations are almost 
simultaneous (ten days apart).

The top panel of Fig. \ref{fig5} shows our NIR echelle spectrum ({\it solid line})
together with a FTS spectrum from \citet{Lambert81} ({\it dotted line}).
The FTS spectrum was obtained during the brightness minimum of 1979 
and reveals the CO band almost completely in absorption.
The echelle spectrum, observed on the descending branch of the 
light curve before brightness minimum, clearly harbors an absorption 
component in CO as well, although the band is dominated by emission in the 
low-resolution spectrum.

The detailed profiles of several CO lines in 2004 are shown in Fig. \ref{fig5}b, 
together with the Na~{\sc i} doublet and Br${\gamma}$ lines.
The prominent Na~{\sc i} emission lines are very symmetric and
centered around the center-of-mass velocity of the star.
The Br${\gamma}$ line is in absorption and its core appears somewhat blue-shifted. 
The broader wings of this strong line extend to $\sim$100 $\rm km\,s^{-1}$ in both 
directions, also observed in H${\alpha}$.
The CO lines however consist of both emission and absorption components,
with fluxes that systematically vary with the excitation energy of each 
transition. The low-excitation CO lines (with small $J$s) are in strong absorption,
while high-excitation lines (with large $J$s around the bandhead) are 
in pure emission. The CO emission line maxima appear to shift redwards with 
decreasing $J$s. We think however that the CO emission lines are 
always centered around the star velocity (or slightly blue-shifted by $\leq$7 
$\rm km\,s^{-1}$) and the progressive emission line redshift away from the bandhead 
is only apparent due to a strengthening of the nearby CO absorption line 
components. With a FWHM of $\lesssim$ 25 $\rm km\,s^{-1}$, the CO emission component is 
narrower than the Na~{\sc i} emission lines and appears round-topped.
The absorption component of CO is centered around 
$V_{\rm rad helio}$ = $-$80 $\rm km\,s^{-1}$ (or $-$33 $\rm km\,s^{-1}$ 
relative to the star velocity), and has a prominent violet line 
wing extending up to at least $-$100 $\rm km\,s^{-1}$ (or $-$53 $\rm km\,s^{-1}$ 
relative to the star velocity).

A detailed comparison of the two CO spectra observed 25 years apart 
reveals a striking transformation from deep absorption lines in 1979
to apparent P Cygni profiles in 2004.
As is indicated by our low-resolution spectra, the transformation 
from absorption to emission actually occurs on a regular basis every 
pulsation cycle. Sheffer's long-term high-resolution NIR 
monitoring of $\rho$ Cas offers a unique opportunity to investigate 
its causes by tracking the detailed evolution of the CO line profiles.
Figure \ref{fig5}c shows a wide diversity in line shapes exhibited by CO in 1979-2004.
For a full set of profiles we refer to Figs. 5.15 and 5.16 of \citet{Sheffer93}.  
When we correlate them with the corresponding pulsation phases and compare 
with optical line profiles, a coherent interpretation can however be proposed.

First, the variations of both emission and absorption line fluxes are 
larger when the amplitude of the light curve ($\Delta$ $V$) is also larger.  
Compare for example the strong CO lines observed in 1979-81 and 2000-04 
with the 1987-89 epoch when the CO absorption lines remain permanently
split. During the latter epoch pulsations are weak (see Fig. \ref{fig4}) and the two 
absorption line components vary only slightly in intensity and Doppler 
shifts. Secondly, prominent CO line emission only occurs near 
phases of maximum light. It is always centered within $\pm$10 $\rm km\,s^{-1}$ 
around the star velocity. On the other hand, the CO line absorption can 
both be either red- or blue-shifted, with shifts of $\lesssim$30 $\rm km\,s^{-1}$.  
The precise dependence of absorption on the corresponding pulsation phase however remains
unclear. This is in part because the CO line absorption obviously becomes  
filled in with variable emission. The outer atmosphere of $\rho$ Cas is also 
velocity stratified and only loosely coupled to the pulsations of its lower 
photosphere, which can yield different Doppler displacements for the CO absorption 
and emission line formation regions. Finally, it is important to point out 
that the properties and the time-scale of these CO line profile variations closely 
resemble the profile evolution observed in optical atomic split lines such 
as the Fe~{\sc i} 5506 \AA\, line in Fig. 2 of \citet{Lobel2003}. 
When one of the split absorption components strengthens the other one weakens.
The double CO core profiles can thus simply be interpreted as one broad
absorption line with semi-regular Doppler displacements, in which core splitting 
occurs when it passes across a static central emission reversal inside the line 
core.

The broad range of NIR CO line morphologies in $\rho$ Cas 
argues firmly against a P Cygni line profile formation region.
The CO emission flux varies on time scales directly linked to 
the variability cycles of the photospheric aborption spectrum. 
This is unlike classical P Cygni profiles observed in ultraviolet 
resonance lines of hot stars that are static and almost 
invariable in time. It signals that the major portion of the CO 
emission line formation region is influenced by the recurrent photospheric 
changes in $\rho$ Cas, rather than for example emission emerging from a very far 
extended optically thin circumstellar envelope that produces redshifted photons 
by backscattering in the far-side expanding hemisphere.
The 2.3 $\mu$m CO emission can therefore not emerge from 
a (possibly) very extended molecular (e.g. CO) gas envelope 
around the supergiant. This is further evidenced by NIR 
CO line profiles observed in HR 8752, another variable cool hypergiant. 
\citet{Sheffer93} observed that either only emission  
occurs in its CO lines, or that it is accompanied by 
\textit{red-}shifted absorption, yielding an {\em inverse P Cygni}-type profile 
instead \citep{Lambert81}. The 2.3 $\mu$m CO feature 
however vanished in HR 8752 in the late 1980s,
approximately at the same time when its optical variability became
much smaller and the effective temperature increased (\S \ref{hyps}). 
We now investigate where the CO emission line formation region
is located in $\rho$ Cas.

\section{Modeling CO Emission}\label{comod}

Since the 2.3 $\mu$m CO band is not observed in 'normal' F-G -type 
supergiants, neither the CO emission nor the absorption components appear to be of 
photospheric origin in $\rho$ Cas. The supergiants of our comparison sample however 
have smaller luminosities with larger gravity,
and molecular bands can be strongly gravity-dependent \citep{Kleinmann86, McGovern04}. 
Furthermore, since $\rho$ Cas is a pulsating star, can the strong variability 
of the underlying photospheric spectrum cause apparent variability in CO, as well as 
in other 'external' spectral lines? The strong correlation of the CO emission 
flux with variability phase signals its formation close to the pulsating photosphere.
We therefore adapt a numerical model of the average photosphere to probe a number of spectral
properties of the CO band observed in the high-resolution CGS4 data.

\subsection{Photospheric Parameters}\label{param}

Every pulsation phase of $\rho$ Cas
requires an estimate of at least two photospheric model parameters:
$T_{\rm eff}$ and log $g$. We first compare the CorMASS spectra of 
$\rho$ Cas with its less-luminous cousins, the Cepheids and the 
non-variable yellow supergiants of our sample.
In the latter stars the photospheric lines are sufficiently narrow 
to apply the classical method of absorption line depth ratios
to reliably infer the $T_{\rm eff}$. Figure \ref{fig1} shows two spectra
of the 3d-period classical Cepheid Y Aur obtained about half a 
pulsation cycle apart. High-resolution optical spectroscopic studies 
show that $T_{\rm eff}$ of short-period Cepheids can vary between 
5500 K and 6500 K with log $g$= 1.9-2.3 \citep{Kovtyukh2000}. 
For the early G non-variable $\mu$ Per in Fig. \ref{fig1} we adopt 
$T_{\rm eff}$ = 5330 K and log $g$ = 1.8 (V. Kovtyukh, private comm.) 
When comparing CorMASS spectra of $\rho$ Cas of 2003 November and 2004 October
to Y Aur and $\mu$ Per we find that $T_{\rm eff}$ should exceed 5500 K for $\rho$ Cas.
To get a better constraint on the temperature, we fit the CorMASS spectra of 
$\rho$ Cas (after dereddening by $A_{V}=1^{\rm m}.3$ \citep{Zsoldos91})
with the Next-Gen -giant model spectra computed with the {\tt Phoenix} 
code \citep{Hauschildt99}. Our best fits confirm a log $g$ value of 
$\sim$0.0, whereas for $T_{\rm eff}$ only a lower limit can be derived 
due to a lack of models with $T_{\rm eff}$ $\ge$ 6600 K and log $g$ $<$ 2.0 
(Fig. \ref{fig6}). We obtain a $T_{\rm eff}$ $\ge$ 6400 K for the 
spectrum of 2003 November, but slightly cooler for 2004 October.
The light curve (Fig. \ref{fig4}) shows that the star assumes maximum 
brightness between 2003 November and 2004 October. The star is dimmer in the 
$V$ band in 2004 October compared to 2003 November, which confirms the smaller 
$T_{\rm eff}$ derived from our best fit procedure. If one assumes that 
the CO absorption component forms in the photosphere and becomes stronger 
with smaller $T_{\rm eff}$, then the shallower CO absorption band observed 
in 2004 October must result from filling in by emission which occurs on 
the descending branch of the light curve. In the following 
Section we investigate whether or not the CO absorption component 
forms in the stellar photosphere.

We can also constrain $T_{\rm eff}$ from our high-resolution 
\textit{optical} spectra. \citet{Lobel2003} established a tight relationship 
between the equivalent width (EW) of the Fe~{\sc i} 5572 \AA\, absorption line 
and the $T_{\rm eff}$ of $\rho$ Cas. Our optical echelle spectrum of 2004 
April 6 indicates an EW for the Fe~{\sc i} line of 348 m\AA\, which corresponds to 
$T_{\rm eff}=$7290 K. In the spectrum of 2004 October 23 we measure an EW$=$465 m\AA\,
yielding a $T_{\rm eff}=$6910 K. In addition, the visual brightness in 2004 October
was comparable to 2003 June, for which Fe~{\sc i} provides $\sim$6800 K.
These estimates agree well with the lower limits estimated from the CorMASS
spectra. They confirm that the hypergiant was cooling down while dimming to 
minimum brightness (after early spring 2004), but it did not cool to below 
6000 K. We can hence adopt a reliable estimate of $T_{\rm eff}=$7200$\pm$100 K for 
the SpeX and CGS4 emission spectra of 2004 September, and of 6800$\pm$100 K for the 
CorMASS absorption spectrum of 2004 October.

\subsection{Temperature Minimum in the Upper Atmosphere?}\label{shock}

We use a Kurucz model of the photosphere with 6750 K and log $g$=0.5 to compute the 
first overtone \textit{absorption} band of CO. We adopt a value of 0.5 
because it is the smallest gravity model available for this temperature.
We perform LTE radiative transfer syntheses of the 2.29-2.32 $\mu$m wavelength
region using Kurucz line lists. The input line data include the 2.3 $\mu$m CO band 
transitions and a number of strong atomic transitions in this NIR region.
The top panel of Fig. \ref{fig7} shows the kinetic gas temperature structure of the model.
The bottom panel plots the computed spectrum ({\em dot-long-dashed line}).
The computed spectrum only shows some weak atomic absorption lines, without 
discernible CO features. We checked that noticeable CO absorption can only be computed
with models of $T_{\rm eff} \le$ 6000 K, significantly below the values we determine 
from our spectral monitoring in 2003-04. These hydrostatic 
models of the photosphere therefore do not produce strong CO absorption bands 
in F-type stars such as $\rho$ Cas. The kinetic gas temperatures in the photosphere 
are too large to yield a sufficient CO abundance, even in the low-gravity 
conditions of super- and hypergiant atmospheres.

Since we observe that the CO emission variability correlates more strongly with the 
stellar pulsations than the CO absorption, we model the emission with atmospheric 
layers in the vicinity of the photosphere. We adapt a Kurucz model for the 2004 
September emission spectrum with $T_{\rm eff}$=7250 K and log $g$=0.5 
(Fig. \ref{fig7}, {\em dot-short-dashed line})
by adding a kinetic gas temperature minimum in the upper photospheric layers over a small
range of optical depths ({\em long-, short-dashed and dotted lines}). 
The modified model produces CO emission when the $T_{\rm g}$-minimum is placed 
below log $m$ $<-4$ in the column mass scale. The observed emission flux at the 
CO bandhead does not appear to be influenced by absorption and can correctly be 
computed if the $T_{\rm g}$-minimum is located around layers with log $m$ = $-$3.1$\pm$0.1. 
The latter column mass value corresponds to only $\sim$5\% of the mean stellar radius 
above the photosphere. We apply a projected microturbulence velocity broadening 
of 11 $\rm km\,s^{-1}$ to compute the NIR spectrum. 
The supersonic microturbulence velocity fits the observed CO emission line 
widths, without invoking extra macroturbulence or rotational line broadening. 
Our composite model produces the observed CO emission flux. 
It however fails to precisely match the detailed shape of the complete 
CO band. For example, the computed spectrum obviously lacks sufficient flux 
longward of the bandhead. Further improvements of the fit would require 
an even more detailed atmospheric model with additional modifications in 
the other thermodynamic conditions besides the $T_{\rm g}$-structure 
(in the current model they are only roughly interpolated to match the
extension of the upper temperature profile). More sophisticated and detailed
non-LTE radiative transfer calculations of CO would be appropriate 
in the very low density conditions of our model, but they are beyond the 
scope of the present investigation.

How can the temperature minimum required in these model fits
be produced in the outer atmosphere of $\rho$ Cas?
A similar atmospheric temperature structure has been proposed for the Sun
to explain the fundamental band of CO at 4.7 $\mu$m observed off-limb
in emission \citep{Ayres02}. The solar CO emission forms in the so-called 
'CO-mosphere', a transitional region located between the photosphere and 
the chromosphere at an altitude of 600-1000 km. 
This intermediate emission region is an inhomogeneous gas mixture with an 
average $T_{\rm g}$ of $\simeq$3000 K permeated by chromospheric 'fingers' of
$T_{\rm g}$ $\sim$10,000 K. There is however no evidence of a classical chromosphere 
in $\rho$ Cas: the most important stellar chromospheric indicators 
such as emission in the cores of the Ca~{\sc ii} H \& K and Mg~{\sc ii} $h$ \& $k$ lines 
have never been observed. 

Because we observe the strong NIR CO emission only during 
variability phases of mean atmospheric expansion it may indicate an alternative 
dynamical mechanism that can cause temporal sharp structures in the 
temperature- and density-profile of the upper atmosphere.
An outwardly propagating field of weak stochastic shock wave trains was
for example proposed by \citet{deJager97} to model the supersonic microturbulence 
velocities observed in the extended atmosphere of $\rho$ Cas.
The CO emission could be produced by a pulsation-driven circumstellar 
shock wave in the cooling layers behind the shock front where the shocked 
gas expands while flowing into the following shock wave.
After shock passage the cool CO layers are heated by compression in the trailing shock
which partially dissociates the molecular CO gas fraction and quenches its local 
emission. 

\section{Discussion}\label{discus}

\subsection{Comparing CO with Peculiar Atomic Lines}\label{atomic}

\subsubsection{Atomic Line Classes}

To compare the NIR CO lines with atomic line profiles, we plot in Fig. \ref{fig8}
representative examples for the main line morphology groups observed in $\rho$ Cas.
The NIR lines are from the UKIRT spectrum of 2004 September 17 with 
emission lines during a phase of declining brightness. The optical lines are 
from the KPNO spectrum of 2004 October 23 when the emission lines weaken before brightness 
minimum. Both spectra have comparable spectral resolution of $\sim$8 $\rm km\,s^{-1}$,
sufficient to resolve individual line profiles. Although the emission spectrum 
weakens in 2004 October (see the CorMASS spectrum in Fig. \ref{fig4}) and the atmospheric 
expansion must have decelerated somewhat, we believe that the overall line shapes 
and Doppler shifts do not appreciably change between the two observations
(less than $\sim$1/12 of $P$ apart). We can therefore directly compare the NIR and optical 
line shapes, within the same pulsation cycle and nearly the same phase,
which is crucial for this long-period semi-regular 
variable star.

In Fig. \ref{fig8} we overplot spectral features of $\rho$ Cas
with high-resolution spectra of Arcturus ($\alpha$ Boo) and the Sun 
(both aligned to the rest velocity of $\rho$ Cas), because their 
spectral lines are narrower and more symmetrical. 
These reference spectra permit accurate line identifications in $\rho$ Cas
whose lines are very broad and often blended or distorted with emission.
Furthermore, the $T_{\rm eff}$ of $\sim$4400 K ($\alpha$ Boo) and 
5800 K (Sun) provide a useful estimate of the
sensitivity of the photospheric absorption spectrum with the $T_{\rm eff}$ 
changes of $\rho$ Cas, which can reveal the intrinsic variability 
of \textit{emission} lines in different pulsation phases.
We compare with Arcturus and the Sun despite the larger atmospheric 
gravity values of log $g$=1.5-2.0 \citep{bell85} and 4.4, respectively, 
than in $\rho$ Cas (log $g$ $\sim$ 0) because detailed synthetic spectra of cool 
hypergiants are presently not available.
Although there is no clear relation between all the spectral lines of
these three cool stars (e.g. strong absorption lines in the Sun and Arcturus
are sometimes absent in $\rho$ Cas), the comparison reveals that the diverse 
line profile shapes in $\rho$ Cas are intrinsic to the peculiar spectrum of 
this hypergiant, and do not result from common absorption line blending.

The optical lines 
of $\rho$ Cas can be divided into four main types we discuss in more detail: 

I. Photospheric absorption lines of neutral and singly ionized metals
(e. g. Fe~{\sc i} 5572 \AA\, used as $T_{\rm eff}$ indicator in \S \ref{param}).
Hydrogen absorption lines can be included in this group as well,
although their line formation regions extend far above the photosphere,
while H${\alpha}$ can exhibit emission components in both line wings
\citep[Fig. 2 of][]{Lobel2003}.

II. `Split' absorption lines with a central emission peak that is essentially constant
in both flux and wavelength. In the stronger lines of multiplet
(with larger oscillator strengths log(gf)) the wings of the emission core
show up against the broad and intensely saturated underlying absorption cores
(example, Fe~{\sc i} 5269 \AA\, line).
The cores of weaker lines (such as Fe~{\sc i} 6191 \AA\,) split only when 
the absorption core crosses over the central emission peak, which occurs 
when the radial velocity of the photosphere is similar to its
mean. The FWHM of unsaturated split lines is comparable to single 
absorption lines ($\sim$60 $\rm km\,s^{-1}$) and indicates a
common photospheric origin for both.  

III. Metallic emission lines of variable strength. They appear prominently above the
level of the stellar continuum soon after maximum brightness, but they diminish and
then completely vanish around minimum light.
A good example of this type of emission line 
is the semi-forbidden Fe~{\sc i} triplet around 0.8 $\mu$m. 

IV. Metallic emission lines of roughly constant strength. This class includes a few
[Ca~{\sc ii}] lines and the 2.2 $\mu$m Na~{\sc i} emission lines.

Figure \ref{fig8} shows a number of illustrative lines for each category.
They represent a range of line strengths and are minimally blended.
The lines progressively change strength from absorption to emission
rather than representing separate line groups. 
The detailed shape of each line appears to depend on the 
excitation energy, and to a lesser degree on the oscillator strength.
Disregarding all hydrogen lines and considering only lines 
with $\chi_{\rm low}$ below a few eV, we find that 
the emission component strengthens while the absorption component weakens 
with decreasing log(gf).
The Na~{\sc i} 2.2 $\mu$m emission lines are the only exceptions because
these moderate excitation lines do not show photospheric absorption in F-G stars
\citep[][our Fig. \ref{fig1}]{Kleinmann86}.

\subsubsection{Putting CO Lines in the Context}

The NIR CO lines contain both emission and absorption components with very 
similar profiles compared to the type II and III atomic lines.
Note however that it has not been established yet if there is any physical 
difference between the type II and
III lines, if they form in two separate 
atmospheric regions or even with different excitation mechanisms.
The morphologic similarity of both line types is apparent with a smooth 
transformation from emission below the local continuum level in the strongest 
split lines to (prominent) emission above the continuum for the weaker 
type III lines (Fig. \ref{fig8}). For example, the central emission cores in 
the split Na $D$ lines can only rise above the continuum level during the 
outbursts of $\rho$ Cas \citep{Lobel2003}. Other examples of these 
mixed line types are presented in Fig. 2 of \citet{Lobel2005}.
Based on the line similarities, \citet{Lobel1997} proposed that emission
components in both types emerge from two circumstellar conic shells  
formed at the interface of a supersonically expanding bipolar wind that collides 
with previously ejected material from the hypergiant's violent 
mass-loss history. At the interface the wind velocity decelerates, 
becomes subsonic, and forms a steady (or standing) shock wave in the 
outer atmosphere. A curve of growth analysis applied to optical multiplets 
of the central emission cores in split Fe~{\sc i} lines yields an excitation 
temperature of 3050 K for the shocked region. Provided that these central 
emission lines are sufficiently optically thin, classical shock wave theory 
yields a thickness for the bipolar conic shells smaller or of the order
of the mean stellar radius (using an average mass-loss rate of 
$\dot{M}$=9$\cdot10^{-5}M_{\odot}$ derived from the far violet extended wings 
of photospheric absorption lines). The observed emission fluxes yield 
distances of 15 to 150 $R_{*}$ from $\rho$ Cas for the location of the 
emitting shock interface \citep[][p.54]{Lobel1997} for realistic cone opening angles 
of more than 30 $\deg$. Smaller opening angles for bipolar outflow would 
provide distances to the shock interface that exceed 
estimates in the literature for the distance to the dust formation 
region near $\rho$ Cas \citep{Jura90}.

At these large distances, strong influences of the pulsating photosphere 
on the emission excitation region are not expected, consistent with the 
constant flux of the central emission cores of split lines (type II). 
The Doppler stability of the emission lines with pulsation phase results 
from the subsonic outflow velocity behind the standing shock wave. 
We however observe different properties for the type III emission lines.
These emission lines can appear sporadically in the core of unsaturated 
absorption lines and their shapes can alter while the emission varies with
the pulsations. It is important to know if these lines are 
intrinsically variable in total flux and Doppler position. 
The top panel of Fig. \ref{fig9} shows the time evolution of the unblended 
emission-free photospheric Fe~{\sc i} 5572 \AA\, line during a complete 
pulsation cycle from 2003 June shortly before minimum light, 
throughout the maximum of 2004 May, until the minimum of 2005 January. 
The outburst spectrum of mid 2000 is also shown. The violet line wing 
develops at the beginning of the brightness decline (the atmospheric 
expansion can be traced by the blue-shifted line centroid) and shrinks
back when the star reaches minimum light. The EW (and central depth) 
of the line increases monotonically from maximum to minimum light
because $T_{\rm eff}$ decreases. The bottom panel of Fig. \ref{fig9} shows a spectral 
region with type III lines in the same pulsation cycle. While the absorption 
spectrum steadily strengthens from 2004 April to December (e.g. the Fe~{\sc i} 5572 \AA\, 
and the long-wavelength wing of Eu~{\sc ii} 6645 \AA\, become stronger), 
the emission component clearly brightens as well between April and July and 
fades by December. The emission line evolution resembles the blue wing behavior 
in the top panel of Fig. \ref{fig9}. The variations of the absorption line wing trace 
changes of the wind opacity which also enhances between April and October 2004. 
The prominent type III emission lines that peak above the continuum  
therefore vary intrinsically. Their variability is linked to the 
photospheric pulsations, since the emission flux increases together 
with the wind opacity during phases of fast atmospheric expansion.

The detailed evolution of the NIR CO resembles more the type III than the type II lines. 
The CO emission flux tightly correlates with the pulsations, signaling it 
is directly influenced by the photospheric dynamics indicated with our models 
in \S \ref{comod}. On the other hand, type II emission lines appear
invariable and must form further away from the surface. There are also 
slight morphological differences between type III/CO and type II emission lines. 
The former lines are centered around stellar rest velocity, while the latter are 
blue-shifted by about $-$5 $\rm km\,s^{-1}$  due to the subsonic outflow behind the 
steady shock interface. The type III and the CO emission lines also appear more 
round-topped compared to the triangular shapes of the strong central emission cores 
in the optical split lines (Fig. \ref{fig8}). The former lines signal a spherically symmetric 
line formation region, while the latter lines can emerge from a more cone-like 
outflow observed nearly pole-on. The differences between the two line groups 
in Doppler shift and line shape are rather subtle, but may be crucial to 
determine whether or not they belong to two separate emission regions around 
$\rho$ Cas. 

The variability of type III and NIR CO emission lines signals formation regions
close to the pulsating photosphere of $\rho$ Cas. But how close?
Our model of the CO emission places the formation region
at only $\sim$5\% of $R_{*}$ because the gas density of the emission region 
decreases rapidly farther above the surface. Figure \ref{fig9} reveals however 
that the {\em optical} emission flux
increases together with the wind opacity determined from the far violet wings of 
the photospheric absorption lines. The properties of the stellar wind dynamics 
are determined by the photospheric pulsations and appear to be linked to the 
emission line formation region as well. If one could observe a time delay between 
the increase of the violet absorption line wings (the mean wind opacity) and 
the onset of the emission line spectrum, it would directly provide the distance the wind 
can travel between the stellar surface (e.g. the base of the stellar wind in the photosphere 
at log($\tau$ $\sim$2/3)) and the actual emission line region. A geometric extension 
of the wind to $\sim$2.5 $R_{*}$ \citep{Lobel98} has been modeled from the photospheric 
absorption lines of $\rho$ Cas, which would disagree with the much smaller distance 
we determine for the NIR CO. Though, part of this difference may arise because our 
models for the CO band emission assume a LTE approximation which breaks down in 
the very extended atmosphere and wind of the hypergiant. If on the other hand
one observes that emission spectrum develops earlier than the violet wings,
then it would favor our shock hypothesis. To discriminate between the
two scenarios, a very precise timing 
of the detailed line profile evolution using more frequent 
spectroscopic monitoring intervals will be required. 

\subsubsection{Na~{\sc i} IR Emission}

We close this section with a discussion of the Na~{\sc i} 2.2 $\mu$m doublet lines.
These NIR lines are always observed in emission in $\rho$ Cas, even during the deep outburst 
minimum of 2000-01 (Fig. \ref{fig3}). 
The two emission lines of the Na~{\sc i} 2.2 $\mu$m doublet 
are of nearly equal strength. On the other hand, the spectra of non-emission supergiants
in Fig. \ref{fig3} show that the short-wavelength absorption line is 
stronger (due to larger log(gf) value and blending with other lines)
than the longer-wavelength one, 
causing the short-wavelength emission component in the $\rho$ Cas UKIRT outburst 
spectrum of 2000 October to appear less conspicuous at the medium resolution.
The NIR Na~{\sc i} emission is also observed in many other cool Ia and Ia0 
supergiants, including HR 8752 
\citep[][our Fig. \ref{fig1}]{Lambert81,McGregor1988,Hrivnak94}. It is
traditionally explained in the literature to result from pumping by stellar 
UV radiation, because of the high excitation energy compared to other
optical emission lines. The excitation energy of the upper level
is 3.8 eV compared to $\sim$2 eV for optical emission lines. 
The upper energy level for the Na $D$ core emission can be populated by 
cascading from the lower level of the NIR Na~{\sc i} doublet through the 
1.138 and 1.140 $\mu$m transitions. Unfortunately, the latter lines fall in a 
wavelength region with strong telluric absorption and cannot reliably
be identified in our SpeX spectrum. Calculations of fluorescence were
originally carried out for another star -- IRC$+$10420 \citep{Thompson77}.
This cool hypergiant has a very large mass loss rate in quiescence (\S \ref{hyps}).
\citet{Thompson77} considered Mg~{\sc i} lines with relatively high 
excitation energy ($\chi_{up}=$ 6 eV) at 1.5 and 1.7 $\mu$m,
and concluded based on their strength that 
they must be optically thick and therefore emerge from a region not far
from the photosphere with sufficiently large electron density, such as a stellar 
chromosphere. There are however no clear indications of a classical chromosphere 
in $\rho$ Cas (\S \ref{shock}). The Mg~{\sc i} lines in the $H$ band of
$\rho$ Cas also appear much weaker compared to Na~{\sc i}, unlike in 
IRC$+$10420 (e.g. compare the 
SpeX spectrum in Fig. \ref{fig2} with Fig. 7 of \citet{Humphreys02}).
It is therefore not clear if one can postulate the existence of a classical 
chromosphere in all cool supergiants with the 2.2 $\mu$m doublet observed in 
emission. The rather triangular shapes of the Na~{\sc i} 2.2 $\mu$m emission lines 
can also point to a line formation region in a collimated outflow, rather than 
in a very extended spherical gas envelope around $\rho$ Cas where the more 
round-topped forbidden [Ca~{\sc ii}] lines form.
Independent of the mechanism producing this permanent and strong line emission,
the Na~{\sc i} 2.2 $\mu$m doublet serves as a powerful indicator for identifying
extreme supergiants in the NIR, even from low-resolution spectra.

\subsection{Comparing CO Emission and Split Lines in $\rho$ Cas 
with Other Cool Luminous Stars}\label{others}

The CO absorption lines we observe in $\rho$ Cas
can originate in the circumstellar gas envelope, whose existence has
been inferred for cooler K-M supergiants from spectral modeling
and interferometric NIR measurements \citep[e.g.,][]{Tsuji88,Tsuji01,Perrin04}). 
Near-IR molecular \textit{emission} bands observed in $\rho$ Cas
are however a much rarer spectral phenomenon. 
The emission excitation mechanism is not likely caused by
photo-ionization from a hot companion star because very long-term
radial velocity monitoring does not reveal any indication of binarity, 
while recent far-UV spectroscopy with $FUSE$ confirms that $\rho$ Cas has 
no significant UV excess either \citep{Lobel2005}, as is for example observed 
for the hot companion in HR 8752 with $IUE$. Variable emission lines 
in cool stars however are often associated with wind variability, 
bulk movements of circumstellar gas due to wind outflow, or to periodically 
shocked layers observed in pulsating stars. CO $\Delta \nu$ 2-0 emission lines are 
observed in low-mass protostars \citep{Biscaya97}, in hot and cool supergiants 
(\citet{McGregor1988a, Kraus00}; \citet{Mozurkewich87, Matsuura02}), 
and in some extreme cases of He-flash objects \citep[like V838 Mon,][]{Rushton05} and 
novae \citep{Ferland79, Rudy03}. The wind and shock in $\rho$ Cas however
must be much more optically thin because of the observed absence of emission 
in the hydrogen lines and of prominent P Cygni-type line profiles (except for H${\alpha}$).

\subsubsection{Cool Hypergiants}\label{hyps}

How does the NIR spectrum of $\rho$ Cas compare to other luminous cool hypergiants 
in the upper HR-diagram? The only two objects that at one time have spectroscopically been
comparable to $\rho$ Cas are the hypergiants IRC$+$10420 and HR 8752. IRC$+$10420 is only 
$\sim$0.1 dex more luminous in log($L$), while HR 8752 is less 
luminous by the same amount than $\rho$ Cas \citep{deJager98}. 
Despite the close similarity of the 
basic stellar parameters, the three hypergiants differ in many other respects. 
IRC$+$10420 has a constant mass loss rate which is an order of 
magnitude larger than the other two,
and is also a maser source of the earliest spectral type known. 
The NIR spectra of 1976 \citep{Thompson77} and 1984 \citep{Fix87} 
revealed Na~{\sc i} emission lines, with the higher members of Brackett series in absorption,
while Br$\gamma$ was not detected. The weakness of the hydrogen lines at 
NIR wavelengths was explained with veiling by the spatially resolved dust envelope.
In 1992 \citet{Oudmaijer94} discovered how blue-shifted emission features
protruded through the NIR hydrogen absorption lines,
completely filling them in by 2000 \citep{Humphreys02}.
Interestingly, a weak CO 2.3 $\mu$m emission band can be present 
in the latter spectrum, although it is observed at noise levels.
The optical spectra show signs of both out- and down-flows.
The variability of this cool hypergiant remains poorly studied. 
\citet{Gottlieb78} present photographic photometry signaling a 
strongly variable object before 1920 which was steadily increasing in
brightness until at least the late 1970s. 
To reconcile all these observations \citet{Humphreys02} suggested that 
the variable spectrum of IRC$+$10420 could entirely originate 
from an opaque circumstellar wind rather than the photosphere, 
mimicking rapid evolutionary changes in the HR-diagram 
near the cool boundary of the Yellow Hypergiant Void.

HR 8752 has a less extreme mass loss rate than IRC$+$10420, making 
it more interesting to compare with $\rho$ Cas. It has its own peculiarities though.
The optical and NIR spectra are long-term variable as in IRC$+$10420.
The spectral changes however have different properties. The emission spectrum has
become systematically weaker while the absorption spectrum and colors indicate 
a steady increase in $T_{\rm eff}$ of $\sim$1000 K per decade over the past 30 years.
The photospheric pulsations have almost ceased while the $V$ amplitude 
decreased from $\pm 0^{\rm m}.2$ in the early 1980s, to $\pm 0^{\rm m}.05$ in the 
mid 1990s \citep{Percy92,Nieuwenhuijzen00}. From 1960 until 1980 it was 
in all aspects very similar to $\rho$ Cas, including periodic line doubling and
the occurrence of emission lines (interestingly also in NIR CO), with both emission 
and absorption components that altered with the pulsation 
phase \citep{Harmer78,Lambert78,Lambert81,Sheffer87,Sheffer93}.

We compare the NIR CO spectra of HR 8752 from \citet{Sheffer93} with the brightness curve 
and observe that the CO emission lines appear at the stellar rest velocity 
while the emission flux peaks around maximum brightness, as we also observe 
in $\rho$ Cas. The CO absorption varies less regularly and may occur both 
red- and blue-shifted. Surprisingly, by 1988 the NIR CO band disappears 
completely, followed by type II and a flux decline in the type III emission lines.
The absorption lines remain however very broad as observed in the optical 
spectrum of HR 8752 in 1998 by \citet{Israelian99}.
We observe no trace of NIR CO in our CorMASS spectra of
2003 November and 2004 October (Fig. \ref{fig1}).
The Na~{\sc i} 2.2 $\mu$m emission lines are still observed, although \citet{Sheffer93}
reported a decrease of the line fluxes. Unlike $\rho$ Cas, 
HR 8752 has a main-sequence early B-type companion that may contribute to 
the hypergiant's emission spectrum. For example, the [N~{\sc ii}] 6548, 6583 \AA\ lines
are thought to originate in a Str\"omgren sphere surrounding the companion, but
we do not observe them in $\rho$ Cas.
What caused the disappearance of CO and other low-excitation lines 
observed before 1987 in HR 8752? We suggest that either the
decrease of the photospheric pulsation amplitude weakened the shock wave(s) 
that can excite these emission lines, or that CO molecules close to 
the photosphere dissociated with the large increase of $T_{\rm eff}$ during 
following decades. But what can cause this very fast increase of $T_{\rm eff}$? 
Possibly we observe this hypergiant in transit across the Yellow Evolutionary Void, 
on its way of becoming a blue supergiant. Or perhaps its atmosphere is warming 
up with the approach of the hot companion. Another possible explanation is 
that HR 8752 exhibits long-term secular variability, perhaps comparable 
to the changes from more to less regular atmospheric oscillations we also 
observe for $\rho$ Cas. Continued spectroscopic and photometric monitoring 
of these unique luminous cool objects is needed to address these questions 
in further detail.

\subsubsection{Post-AGB Supergiants}

Hypergiants are very exceptional luminous stars. 
Further insights of their NIR spectroscopic variability can be gleaned 
from more abundant and more regularly pulsating less-luminous cool supergiants.
Good examples of stars for which we also observe NIR CO in emission are a 
number of yellow post-AGB stars and some proto-planetary nebulae precursors. 
It is unclear at present if NIR CO emission occurs more frequently in 
variable stars or for a certain spectral type or subclass.
\citet{Hrivnak94} found three proto-planetary nebular supergiants with 
CO emission. The earliest is G0 Ia IRAS 22223+4327, a pulsating variable.
\citet{McGregor1988} found CO emission in a few early-type 
Ia-supergiants in the Magellanic Clouds, but they are 
not known to be photometrically variable. The earliest reported case is the 
RV Tau F4-type Ibp star AC Her (HD170756), which
also shows Br${\gamma}$ in absorption and Na~{\sc i} 2.2 $\mu$m in emission
similar to $\rho$ Cas. NIR CO in this star switched from emission to absorption
in three low-resolution spectra of \citet{Oudmaijer95}. We have reduced an archival 
CGS4 spectrum of another post-AGB Srd variable star  
with NIR CO emission, HD 179821 (G5 Ia). The spectrum was obtained on the same night 
of 2000 October 18 as the outburst spectrum of $\rho$ Cas.
It shows the Na~{\sc i} 2.2 $\mu$m lines in emission
but not a trace of CO, confirming the CO variability
previously noticed by \citet{Oudmaijer95}.

\subsubsection{R CrB Stars and $\rho$ Cas Outbursts}\label{RCrB}

The NIR CO bands observed in R CrB stars offer an interesting comparison to $\rho$ Cas.
These less-luminous eruptive variables have carbon-rich and hydrogen-deficient atmospheres, 
but exhibit many similarities to $\rho$ Cas otherwise. They cover the same
range of $T_{\rm eff}$, reveal prominent emission lines at times, and are known pulsators 
that most importantly exhibit deep visual brightness declines \citep{Clayton96}. 
It is thought that occultations by dense dusty circumstellar clouds cause the abrupt 
dimming. The dust clouds should occur rather close to the photosphere, dictated 
by the short time scales of the sudden brightness decreases. But how can dust form 
at photospheric gas temperatures above 5000 K in these F--G type stars?
\citet{Hinkle95} observed the CO 2.3 $\mu$m band in FG Sge, a post-AGB and 
He-flash supergiant, two months before it went into a deep brightness decline,
which turned it into a typical R CrB star thereafter.
They observe split CO line profiles with one absorption component near
the stellar rest velocity and the other blue-shifted by 21 $\rm km\,s^{-1}$.
The CO lines are interpreted as evidence of a long-sought reservoir of 
cool material forming near the stellar surface. After the first circumstellar
gas shell forms at a distance of $\sim$2 $R_{*}$, the second shell 
cools below 1000 K and forms dust while becoming accelerated outwards 
by radiation pressure on the dust. Based on these observations 
\citet{Gonzalez98} proposed that CO gas might be the leading coolant  
initiating the nucleation of dust grains very close to the photosphere. 
However, a very recent extended study of R CrB stars by \citet{Tenenbaum05} 
did not establish any clear correlation between the appearance of the 
2.3 $\mu$m CO band and phases of large brightness decrease. 
In fact, they model the observed CO band strengths with pure photospheric 
models. Their work points to a strong connection with $\rho$ Cas for which
we observe that the absorption line splitting is due to a pulsation 
phase dependent superposition of absorption and emission line components that 
recur on a regular basis, rather than being direct tracers of cool 
circumstellar material expelled during outbursts.  

There are convincing indications that the trigger for dust 
formation is actually located in the photosphere of R CrB stars. 
\citet{Gonzalez98} found that the photosphere cools by at least 
$\sim$1000 K during the fast brightness declines. This atmospheric 
phenomenon is often neglected because of difficulties in interpreting
R CrB spectra observed at brightness minima. The absorption lines appear 
to weaken and to become distorted, possibly because the photospheric spectrum
is blurred by the scattering from the expanding dust clouds (although these effects 
are not clearly observed in $\rho$ Cas). The atmospheric cooling itself appears 
to result from some kind of dynamic transformation or instability.
\citet{Rao99} (see their Fig. 1) observe that at the beginning of a strong 
brightness decrease in R CrB the low-excitation photospheric lines reveal an emission 
component and become strongly red-shifted. The high-excitation lines in contrast 
become blue-shifted. This signals that the upper atmosphere decouples from
the regular pulsation movements. The photosphere collapses, causing a 
stronger than usual compression, which in turn results 
in a stronger than usual atmospheric expansion with a large brightness decrease.
A very similar behavior is observed during the 2000-01 outburst of $\rho$ Cas. 
The event was preceded by a strong collapse of the atmosphere which reversed
to expansion and subsequently cooled the entire atmosphere down by more than 3000 K.
The larger brightness decreases in R CrBs compared to the outbursts of 
$\rho$ Cas are likely due to more vigorous dust formation around the former stars. 
The enhanced dust formation may be caused by the large carbon abundance and hydrogen 
deficiency of R CrB atmospheres. Whether or not there is dust formation 
during the outbursts of $\rho$ Cas, it appears reasonable to attribute the 
sudden eruptions to a common physical mechanism operating in the atmospheres 
of cool hypergiants and R CrBs. \citet{Lobel2003} showed that the
recombination of partially ionized hydrogen in $\rho$ Cas, and the recombination of ionized helium
in R CrBs can effectively drive a runaway explosion over the time scales corresponding 
to the observed brightness declines of both eruptive variables. The eruptions are 
triggered by unusual thermal conditions in F-type stars during phases of very 
strong atmospheric compression causing a dynamically unstable atmosphere 
\citep{Lobel2001}. \citet{Lobel2003} proposed that the stellar wind and oscillations 
can suddenly activate the trigger, which we observe is tightly correlated 
with the occurrence of strong emission line spectra. The study of strong 
transient emission lines in different types of luminous pulsating cool stars 
therefore provides important clues to the physical mechanisms that produce 
the large diversity of their observed light curves. 

R CrB stars both in and out of the sudden brightness declines reveal many 
spectral similarities with $\rho$ Cas. 
Most noticeably the photospheric lines of R CrBs are very broad as in 
F-type hypergiants \citep{Pandey04}. The lines sometimes split 
around phases of maximum brightness, which appears to correlate with the 
radial velocity amplitude. In RY Sgr, having a pulsation amplitude 
$\pm$20 $\rm km\,s^{-1}$, the split lines occur regularly, while in 
R CrB, having amplitude of only $\pm$3 $\rm km\,s^{-1}$, 
they are not observed every pulsation cycle, except in the strongest lines \citep{Rao97}.
$\rho$ Cas has an intermediate velocity amplitude of $\pm$8 $\rm km\,s^{-1}$, 
between these two values. The brightness declines are preceded by an increase 
of the pulsation amplitude, observed for example in R CrB \citep{Rao99}, and in 
$\rho$ Cas by \citet{Lobel2003}. Possibly, it is therfore no surprise that the 
emission line spectrum also strengthens during variability phases preceding the 
outbursts of R CrBs. The so-called E1 group in R CrBs represents transient sharp emission 
lines that are only briefly observed at the very onset of the declines.
The E1 group may therefore correspond to our variable type III emission lines in $\rho$ Cas.
On the other hand, the so-called E2 group of emission lines
has spectral characteristics identical to our type II lines.
These emission lines are sharp cores inside low-excitation absorption lines,
and are permanently blue-shifted by 3-5 $\rm km\,s^{-1}$ relative to the star velocity.
They are apparently of constant flux and appear most prominently in the middle of 
the fast brightness declines when the stellar continuum flux significantly decreases
\citep{Cottrell82, Gonzalez98, Rao99, Skuljan02}. 
Sharp emission line components above the local continuum level are observed 
during brightness minimum in R CrB stars in forbidden lines like
[Ca~{\sc ii}] and in resonance doublet lines of K~{\sc i} (0.77 $\mu$m) 
and Na $D$ (0.59 $\mu$m). The narrow emission components however do not occur in Mg~{\sc ii} 
$h$ \& $k$ lines (at 0.28$\mu$m) or Ca~{\sc ii} H \& K (0.39 $\mu$m) \citep{Clayton94, Rao99}, 
which again compares to the observations of $\rho$ Cas. 
Near-IR Na~{\sc i} or CO emission lines have never been reported in R CrB stars,
possibly because this spectral region has not thoroughly been investigated.
The optical $\rm C_{2}$ bands however are observed to appear in emission 
at the onset of the brightness declines. They either remain in emission or return to 
absorption around minimum brightness, although a clear correlation with the pulsation phase 
or phases of brightness declines has yet to be established \citep{Gonzalez98, Rao99, Skuljan02}.
The two most significant spectroscopic \textit{differences} between R CrB stars
and $\rho$ Cas are the presence in the spectrum of the former and the absence in the 
latter of: 
1) a strong He~{\sc i} 1.083 $\mu$m line with a P Cygni profile (although with notable 
exceptions for warmer $T_{\rm eff}$ $>$ 6000 K stars as V854 Cen),
indicative of warm chromospheric winds in R CrB stars \citep{Clayton03}, and of
2) a broad (FWHM $>$200 $\rm km\,s^{-1}$) emission component observed during 
R CrB brightness declines in forbidden and permitted strong lines (such as 
H$\alpha$, He, and metal resonance lines), which \citet{Rao99} suggest can 
result from an accretion disk around a putative white dwarf companion star 
(but note that a hot companion is not observed in $\rho$ Cas). 

\subsubsection{Line Splitting and Pulsations}

There is ample evidence from the literature that other long-period 
cool variables such as Cepheids, RV Tau stars, R CrBs, and Miras show absorption line 
splitting and emission lines above the continuum level around phases of maximum brightness,
in atomic as well as molecular lines of CO and CN \citep{Sanford52,Kovtyukh03,
Alvarez01,Mozurkewich87,Rao97,Nowotny05b}.
The line splitting phenomenon is also observed in hot RR Lyr variables 
\citep{Chadid96}. Based on the above discussion we think it is 
merely a matter of coordinating coeval optical and NIR 
observations to firmly establish that phase-dependent spectral changes in 
absorption line splitting and emission line flux are related spectroscopic 
phenomena, because spectral variability is clearly linked with changes of 
$T_{\rm eff}$ in cool pulsating stars. These spectral phenomena are wide-spread 
among pulsating stars of various types in the cool part of the HR-diagram. 
They should therefore be attributed to radiative transfer effects that result 
from atmospheric pulsations instead of expanding circumstellar gas shells that 
are simply postulated by default for detailed calculations of mass-loss rates.

\subsection{Emission by the Running Shock Wave?}

Considerable theoretical work has been published to model emission 
lines in \textit{cool} pulsating stars. Among all cool stars with split absorption lines, 
the Mira variables are possibly best studied with recent dynamic models of the 
atmosphere that quite well match the observed line profiles.
In most studies however the apparent absorption components of split absorption   
lines are still interpreted to result from separate detached layers (optically thick 
'shells') with different Doppler velocities. The double line cores are simply assumed to 
have a purely kinematic origin \citep{Bessel96, Winters00, Nowotny05a, Nowotny05b}.
The alternative interpretation we propose with a central emission reversal inside 
the absorption line core is computationally much more challenging and has only 
very recently been explored with radiative transfer calculations.
\citet{Richter03} consider specific spectroscopic examples with Fe~{\sc ii} and 
[Fe~{\sc ii}] emission lines that occur in some Miras only after exceptionally 
large brightness maxima (or equivalently very strong atmospheric compression).
They demonstrate that these emission lines can be excited in a confined 
post-shock layer close to the stellar photosphere at 1-2 $R_{*}$. The post-shock 
gas is heated by the passage of a shock wave produced in the photosphere, 
which reaches this emission layer in half a pulsation cycle. For the shock to be 
sufficiently strong to produce emission lines, a minimum shock velocity of 
$\sim$20 $\rm km\,s^{-1}$ is required in Miras. In an earlier study \citet{Woitke96} 
already demonstrate that adiabatic cooling of shocked gas can yield sufficiently 
small gas temperatures for the nucleation of dust grains within the same close 
distance of 1.5-3 $R_{*}$ from photosphere of R CrB stars. 
The local gas density and shock velocity are the prime factors that determine the 
location of the emission zone. These models support the small distances of only 
$\lesssim$1 $R_{*}$ we compute for the NIR CO emission line formation region in $\rho$ Cas 
with a simplified atmosphere model in which the outer gas temperature structure is 
modified by a shock wave.

Some difficulties however remain when trying to directly apply the above models
to the warm yellow hypergiants $\rho$ Cas and HR 8752.
It should be remembered that radiation pressure on dust around the dust condensation 
radius (at a few $R_{*}$) is considered to be the driving mechanism of winds and mass loss 
in all modern models of R CrB and Mira variables. In $\rho$ Cas however, there is no 
evidence so far of significant dust nucleation close to the photosphere. The weak 9.8 
$\mu$m silicate dust emission feature and the IR spectral energy distribution place 
the dust envelope (possibly resulting from the outburst of 1945-46) at 30-130 $R_{*}$ from 
the star \citep{Jura90}. New IR photometry and spectroscopy after the 2000-01 outburst 
would help to further determine the properties of circumstellar dust in $\rho$ Cas. 

Furthermore, some constraints from our NIR CO emission models can be used to
estimate the velocity of the shock wave near $\rho$ Cas.
When index 1 denotes the pre-shock (upstream) gas conditions $T_{\rm 1}$, $p_{\rm 1}$, 
and $\rho_{1}$, and index 2 the post-shock (downstream) conditions (with jump
conditions across a compression shock surface:
$T_{\rm 2} > T_{\rm 1} $, $p_{\rm 2} > p_{\rm 1}$, and $\rho_{2} > \rho_{1}$), the 
speed of the shock $u_{1}$ can be expressed through the local adiabatic speed of sound 
and the Mach number: $u_{1}=c_{1} \cdot M_{1}$. For a specific heat ratio of ideal 
monatomic gas of $\gamma = 5/3$ (assuming hydrogen gas) to $7/5$ (for CO gas with rotational 
degrees of freedom) and the mean gas-pressure and -density structures of our 
atmosphere model (\S \ref{shock}), we compute $c_{1}$=$\sqrt{\gamma \cdot p_{1} / \rho_{1}}$ = 7.5-7  
$\rm km\,s^{-1}$. The pre-shock Mach number $M_{1}$ can be computed with the 
classical Rankine-Hugoniot jump relations in polytropic gas which are a function 
of $\gamma$ and $T_{2}/T_{1}$ only. For $T_{2} / T_{1} = 4200/2500$ K 
(Fig. \ref{fig7}) we compute $M_{1}$ = 1.30-1.36. We thus obtain a shock velocity of 
$u_{1}$ $\sim$10 $\rm km\,s^{-1}$. This value is smaller than the 20 $\rm km\,s^{-1}$ 
threshold velocity required in the shock models to excite atomic emission lines, 
and yet they are observed every pulsation cycle in $\rho$ Cas. 10 $\rm km\,s^{-1}$ is 
in fact only a lower limit, since $T_{2}$ (the gas temperature of the post-shock 
region) can be much larger than 4200 K we adopt from the unperturbed model atmosphere.
On the other hand, the non-detection of dust formation in the vicinity of the photosphere 
of $\rho$ Cas appears consistent with the small velocity (or Mach number) of the shock 
wave(s). 

\citet{Alvarez01} conducted a large survey of the absorption line doubling 
phenomenon in long-period variables. They demonstrate that it is more frequently 
observed among stars with compact rather than extended atmospheres, 
and interpret it to result from stronger shock waves in the former stars.
Hypergiants such as $\rho$ Cas and HR 8752 with their enormous stellar 
radii of several hundreds of R$_{\odot}$  do clearly not follow this trend.
On the contrary, HR 8752 has become more compact as $T_{\rm eff}$ has
dramatically increased over the past 30 years, but the split absorption
lines are presently no longer observed in its optical spectrum compared to 
the 1970s. If the line broadening of these split lines were solely due to 
a superposition of two absorption components emerging from two separate gas layers 
above and below the shock surface, predicted with models of Miras, the 
width of these lines would have to decrease with the disappearance of the shock 
waves in the atmosphere of HR 8752. This is however not observed because the 
photospheric lines typical of a hypergiant spectrum have stayed very broad. 

Our model for the central emission excited by a propagating shock wave
possesses some difficulties as well. A shock wave propagating supersonically 
in the stellar rest frame should produce red-shifted CO line emission 
(provided the emission is sufficiently optically thick), which we do not observe. 
Another problem is that a transient and spatially rather narrow atmospheric 
structure required for the CO emission cannot readily account for the permanent 
excitation of other static (and apparently centered around stellar rest) atomic 
emission lines in our high resolution spectra.

In summary, none of the available theoretical models of the atmospheric dynamics 
can coherently explain the absorption line splitting and line emission phenomenon 
in pulsating cool supergiants. The field of research is making rapid progress 
however. The current investigation shows that the correct answer may be obtained
with the development of a new generation of atmospheric models that couple
the dynamics of the atmosphere with non-LTE excitation of spectral lines in 
shock waves. 

\section{Summary and Conclusions}\label{concl}

We describe a high-resolution spectroscopic study of one of the most 
luminous cool stars in the Galaxy, the bright yellow hypergiant $\rho$ Cas.
Our observations cover a complete pulsation cycle in 2003-04 over a large 
wavelength range from the optical to $\sim$4 $\mu$m. Near-IR spectroscopic 
studies of yellow supergiants are scarce, and we demonstrate that the emission 
lines they harbor serve as powerful diagnostics of the important atmospheric 
processes of mass loss and pulsation in massive stars. In particular, we 
identify several prominent emission lines in the $L$ band spectrum observed 
near maximum brightness, and present $K$-band spectra obtained during the 
last grand outburst of the enigmatic hypergiant in 2000-01.

We primarily investigate the first overtone band of CO at 2.3 $\mu$m because it 
occurs prominently in emission during certain variability phases of $\rho$ Cas.
The NIR CO band is believed to originate from expanding shells and to trigger 
the formation of dust close to the photosphere of R CrB stars. 
We confirm the findings of \citet{Sheffer93} that the CO band in $\rho$ Cas 
is variable and consists of absorption and emission line components. 
However, based on detailed comparisons with atomic lines and correlating 
all spectroscopic observations with the corresponding pulsation phases, 
we argue that the `split' absorption line profiles are not caused by two 
separately ejected shells during rare outbursts of the hypergiant, but rather 
are a commonly observed combination of a static narrow central emission line 
superimposed on the core of a broad absorption line.
The absorption line exhibits Doppler displacements due to the atmospheric 
pulsations, while the emission component is always centered around the stellar
rest velocity. The flux of the central emission line varies with the  
pulsations of the photosphere and rises above the continuum level around 
maximum brightness during phases of fast atmospheric expansion. 
After the phase of minimum brightness the emission flux decreases and 
the lines fade. Detailed comparisons of NIR 
CO line profiles with low-excitation atomic lines reveal that they are 
intermediate between the type II and type III lines of $\rho$ Cas 
(according our terminology). While CO absorption components are always 
present and vary similarly to permanently split atomic absorption lines 
(type II), the large variability of the superimposed CO emission components 
suggests they are more compatible with type III emission lines that can 
appear above the level of the stellar continuum flux. In other words, the
excitation mechanism of prominent atomic emission lines and CO emission 
we observe in $\rho$ Cas can have a common physical origin.

The strong correlation of the NIR CO emission flux we observe 
with the radial velocity curve from photospheric pulsations signals that 
the CO emission lines form in the vicinity of the photosphere. 
We model the CO emission band spectra by inserting  a cool gas layer 
of $T_{\rm g}$ $\sim$2500 K in the kinetic gas temperature structure 
of a plane-parallel hydrostatic Kurucz model of the photosphere with 
$T_{\rm eff}$ =7250 K and log $g$ =0.5. We fit the observed emission 
flux around the CO bandhead, which is least influenced by absorption, 
with radiative transfer calculations to constrain the optical depth of 
the cool CO layer. We determine an atmospheric column mass along the 
line of sight of 10$^{-3.1 \pm 0.1}$ g cm$^{-2}$. It translates to a 
geometric distance for the cool CO layer of only $\sim$5\% of $R_{*}$ 
above the photosphere. 

We further compare the evolution of NIR emission lines in $\rho$ Cas with
the spectra of other late-type supergiants. We consider cool Miras, the yellow 
hypergiants HR 8752 and IRC$+$10420, some less-luminous post-AGB stars, 
and their population II analogs of R CrB stars with large visual 
brightness declines. We observe a number of important spectroscopic similarities 
between all these cool variable star types. The most remarkable are: 
1) when several epochs of observations are available the NIR CO emission flux 
is always observed to vary; 
2) the CO emission flux is largest around 
maximum brightness, also observed in prominent atomic emission lines;
3) prominent emission lines and split absorption lines are more frequently 
observed in stars with larger pulsation amplitudes or with larger radial velocity 
amplitudes;
4) the onset of (sudden) strong brightness decreases in $\rho$ Cas and R CrB stars
results from a stronger than usual compression of the photosphere 
followed by a stronger than usual expansion and cooling of the entire atmosphere
with the appearance of the type III variable emission lines we propose are analogous 
to type E1 emission lines observed in R CrB stars;
5) during the fast brightness declines of outbursts the initial emission line spectrum 
is replaced by another type of emission line, the so-called E2 lines in R CrB stars,
and in $\rho$ Cas the type II emission line cores of split absorption lines.
We propose that the latter types of emission lines are always present but become 
more obvious during fast brightness declines because the mean photospheric flux 
suddenly decreases.

In summary, we identify two groups of narrow emission lines with FWHM $<$ 
50 $\rm km\,s^{-1}$ that coexist in the spectra of low-gravity pulsating stars.
The first group consists of atomic lines that appear to be 
intrinsically invariable (types II and IV in our terminology). 
The second group consists of atomic and CO emission lines that are intrinsically 
variable (type III). 
Both groups are low-excitation lines with 
$\chi_{\rm low}$ of the order of few eV having Doppler velocities of 
$\sim$5 $\rm km\,s^{-1}$ around the stellar rest velocity. The excitation 
of both groups appears to be related in one way or another to the atmospheric 
pulsations. The best interpretation we can propose based on our spectroscopic 
observations is that the first group is excited in two static conic regions 
$\sim$10-100 $R_{*}$ above the photosphere \citep{Lobel2003}, where the 
supersonic wind of $\rho$ Cas collides with circumstellar material
(and possibly replenished by recurrent outbursts) forming a permanently 
standing shock wave.
The second group of emission lines is excited in a short-lived circumstellar 
cool gas layer in the immediate vicinity of the photosphere less than 1 $R_{*}$ 
away, that is excited by the passage of a pulsation-driven shock wave.

The following list of observations would be helpful to further verify our interpretations:
\begin{enumerate}
\item  very high-resolution spectroscopic observations ($R$ $\gtrsim$ 100,000) of 
type II and III emission lines in $\rho$ Cas to search for possible 
differences in radial velocity and line profile shapes between both groups.
The observations could confirm if the emission lines form in two spatially 
separate regions in the circumstellar environment of the hypergiant;  
\item high-resolution spectroscopic observations beyond 3 $\mu$m to identify and 
investigate the properties of other prominent emission lines expected in this 
wavelength region;
\item monitoring of the \textit{fundamental} band of CO at 4.6 $\mu$m to determine 
how far out the photospheric pulsations and shock waves propagate into the 
circumstellar environment;
\item new observations of the silicate 9.8 $\mu$m dust emission feature 
after the outburst of 2000-01. Near- and far-IR photometry can establish if 
$\rho$ Cas produces dust during the large brightness decreases of outbursts. 
It can further link its remarkable atmospheric dynamics to that of 
other eruptive variable stars;
\item high-resolution spectroscopic monitoring of NIR CO and split absorption lines 
in other late-type stars. In particular, observations of semi-regular pulsating 
variables (e.g. RV Tau, R CrB stars) can test the relative importance of the pulsations versus
$T_{\rm eff}$ for sustaining the cool gas layers that produce the peculiar emission line 
spectra.
\end{enumerate}

\acknowledgments

We would like to thank all the observers who helped to obtain the NIR spectra.
We also thank staff at Vatican Obs., Steward Obs., and the CorMASS group for providing 
observing time and support during NG's visits to VATT.
The Service Mode Observing group at UKIRT is thanked for their timely observations, 
help with retrieving archival data, and advice on the data reduction.
We thank the AAVSO international database maintainers, with contributions from 
many observers worldwide to the light curve of $\rho$ Cas. Special thanks 
to K. Hinkle for drawing attention to the FTS KPNO spectra of $\rho$ Cas 
and HR 8752. J. Muzerolle and the anonymous referee are 
acknowledged for helpful comments on the manuscript. 
AL acknowledges partial financial support from {\em FUSE} grants GI-D107 and 
GI-E068 by the Johns Hopkins University.

\clearpage  
\begin{deluxetable}{lrcccrrrrc}
\tablecolumns{10} 
\tablewidth{0pt}
\tabletypesize{\tiny}
\tablecaption{Log of spectroscopic observations\label{table1}.}
\tablehead{ 
\multicolumn{1}{c}{Obs. Date} & \multicolumn{1}{c}{Julian Date} & \multicolumn{1}{c}{Telescope-Instrument} &
\multicolumn{1}{c}{Grating} & \colhead{Wavelength range\tablenotemark{a}} & \multicolumn{2}{c}{Resolution} &
\multicolumn{2}{c}{Telluric standard} & \multicolumn{1}{c}{Observers\tablenotemark{b}} \\
\colhead{} & \colhead{2,400,000+} & \colhead{} & \colhead{} & \colhead{$\mu$m} & \colhead{$\lambda / \Delta \lambda$} & \colhead{km/s}  & 
\colhead{HD $\#$} & \colhead{SpT} & \colhead{} 
}
\startdata
1998 Oct 21 &   51108  &UKIRT-CGS4     &40 l/mm        & 1.820-2.450   & 850     & 350           & 195295 & F5Iab   &J. B., G. F.\\
\\
2000 Jul 19 &   51744  &WHT-UES        &E31            & 0.4050-0.8800   & 50,000  & 6         &    --  &   --    &A. L.       \\                                                                                                                       
\\
2000 Oct 18 &   51836  &UKIRT-CGS4     &150 l/mm       & 2.040-2.114   & 6,400   & 50            & 221756 & A1 III  &C. D.       \\
2000 Oct 18 &   51836  &UKIRT-CGS4     &150 l/mm       & 2.115-2.194   & 6,700   & 45            & 221756 & A1 III  &C. D.       \\
2000 Oct 18 &   51836  &UKIRT-CGS4     &150 l/mm       & 2.192-2.353   & 3,500   & 85            & 170296 & A1 IV/V &C. D.       \\
2000 Oct 19 &   51837  &UKIRT-CGS4     &150 l/mm       & 2.035-2.112   & 6,400   & 50            & 221756 & A1 III  &C. D.       \\
2000 Oct 19 &   51837  &UKIRT-CGS4     &150 l/mm       & 2.113-2.192   & 6,700   & 45            & 221756 & A1 III  &C. D.       \\
2000 Oct 19 &   51837  &UKIRT-CGS4     &150 l/mm       & 2.189-2.352   & 3,500   & 85            & 221756 & A1 III  &C. D.       \\
\\  
2003 Jun 10 &   52800  &NOT-Sofin      &Cam2           & 0.4251-0.7744 & 80,000  & 4        &    --  &   --    &A. L., I. I. \\
\\
2003 Nov 20 &   52964  &VATT-CorMASS   &40 l/mm        & 0.650-2.500   & 300     & 1,000         &   4614 & G0 V    &N. G., N. W.\\
\\
2004 Apr 06 &   53101  &NOT-Sofin      &Cam2           & 0.4170-0.9690   & 80,000 & 4       &    --  &   --    &A. L., I. I. \\
\\
2004 Jul 29 &   53217  &NOT-Sofin      &Cam2           & 0.3465-1.024    & 80,000  & 4         &    --  &   --    &A. L., I. I.\\
\\
2004 Sep 6  &   53255  &IRTF-SpeX      &ShortXD        & 0.805-2.400   & 2,000   & 150           & 205314 & A0V     &A. B.       \\
2004 Sep 6  &   53255  &IRTF-SpeX      &LongXD1.9      & 1.910-4.104   & 2,500   & 120           & 223386 & A0V     &A. B.       \\
\\
2004 Sep 16 &   53265  &UKIRT-CGS4     &40 l/mm        & 1.870-2.504   & 900     & 350           & 219290 & A0 V    &S. L.       \\
2004 Sep 17 &   53266  &UKIRT-CGS4     &Echelle        & 2.1584-2.1729 & 37,000  & 8             & 219290 & A0 V    &C. D.       \\
2004 Sep 17 &   53266  &UKIRT-CGS4     &Echelle        & 2.1994-2.2131 & 37,000  & 8             & 219290 & A0 V    &C. D.       \\
2004 Sep 17 &   53266  &UKIRT-CGS4     &Echelle        & 2.2893-2.3034 & 37,000  & 8             & 219290 & A0 V    &C. D.       \\
2004 Sep 17 &   53266  &UKIRT-CGS4     &Echelle        & 2.2994-2.3134 & 37,000  & 8             & 219290 & A0 V    &C. D.       \\
2004 Sep 17 &   53266  &UKIRT-CGS4     &Echelle        & 2.3094-2.3232 & 37,000  & 8             & 219290 & A0 V    &C. D.       \\
\\
2004 Oct 23 &   53302  &Mayall-ECHLR   &Echelle        & 0.5255-0.8523 & 42,000  & 7         &   --   & --      &J. S.        \\
\\ 
2004 Oct 30 &   53309  &VATT-CorMASS   & 40 l/mm       & 0.650-2.500   & 300     & 1,000         &   4614 & G0 V    &N. G., J. G.\\
2004 Oct 31 &   53310  &VATT-CorMASS   & 40 l/mm       & 0.650-2.500   & 300     & 1,000         &   4614 & G0 V    &N. G., J. G.\\
\\ 
2004 Nov 27 &   53337  &NOT-Sofin      &Cam2           & 0.3545-0.9875 & 80,000  & 4       &    --  &   --    &A. L., I. I. \\
\\
2004 Dec 25 &   53365  &NOT-Sofin      &Cam2           & 0.3580-1.0325 & 80,000  & 4       &    --  &   --    &A. L., I. I. \\
\enddata
\tablenotetext{a}{Air wavelengths, as observed.}
\tablenotetext{b}{J. B. -- J. Buckle, G. F. -- G. Fuller, A. L. -- A. Lobel, C. D. -- C. Davis,  I. I. -- I. Ilyin, N. G. -- N. Gorlova,
N. W. -- N. Woolf, A. B. -- A. Burgasser,  S. L. -- S. Leggett, J. S. -- J. Stauffer, J. G. -- J. Greissl }
\end{deluxetable} 

\begin{deluxetable}{rrrr}
\tablecolumns{4} 
\tablewidth{0pt}
\tabletypesize{\small}
\tablecaption{Unindentified emission lines in the $L$ band spectrum of $\rho$ Cas.\label{table2}}
\tablehead{ 
\colhead{$\#$} & \colhead{$\lambda$\tablenotemark{a}} & \colhead{FWHM} & \colhead{EW} \\
\colhead{} & \colhead{($\mu$m) } & \colhead{($\mu$m) } & \colhead{($\mu$m) } \\
}
\startdata
1 & 3.6584 ($\pm$ 0.0015) & 0.0016 & 0.00005 \\
2 & 3.6991 ($\pm$ 0.0015) & 0.0022 & 0.00009 \\
3 & 3.7456 ($\pm$ 0.0015) & 0.0013 & 0.00012 \\
4 & 3.9115 ($\pm$ 0.0016) & 0.0016 & 0.00008 \\
5 & 3.9885 ($\pm$ 0.0016) & 0.0023 & 0.00011 \\
6 & 4.0521 ($\pm$ 0.0016) & 0.0013 & 0.00007 \\
7 & 4.0609 ($\pm$ 0.0016) & 0.0017 & 0.00004 \\
\enddata
\tablenotetext{a}{Rest wavelength in air -- observed wavelength corrected for the Earth's motion
and for the heliocentric radial velocity of $\rho$ Cas ($-$47$\pm$1 $\rm km\,s^{-1}$); uncertainties in brackets --
width of the resolution element.}
\end{deluxetable} 

\clearpage

\begin{figure}
\caption{CorMASS $R$ $\sim$300 spectra of $\rho$ Cas compared to other yellow 
hyper- and supergiant stars. The spectra have been normalized at 1.6 $\mu$m 
and shifted vertically for clarity. Clipped out are the noisy regions
due to strong telluric absorption. Identified are the most prominent absorption 
and emission lines. The inset panel shows an expanded view of the $K$ band spectrum.  
Notice that the NIR CO band is only present in $\rho$ Cas. The line identifications are 
based on high-resolution atlases of the Sun and Arcturus \citep{atlas}.
}\label{fig1}
\end{figure} 

\begin{figure}  
\caption{SpeX $R$ $\sim$2,000 continuum normalized emission spectrum of $\rho$ Cas. Marked are some
of the prominent emission lines previously reported in the literature, as well as 
some unidentified lines in the $L$ band discovered in this work. 
The strong absorption lines beyond 1.5 $\mu$m are mainly due to hydrogen.
}\label{fig2}
\end{figure}

\begin{figure} 
\caption{CGS4 spectra of $\rho$ Cas observed on  two consecutive nights in the 
middle of the deep brightness decrease of 2000-01, compared to other bright 
supergiants ordered according to spectral types, and to the emission spectrum
of 2004. The comparison spectra are offered 
in \citet{Wallace97}. All spectra have been convolved with the same instrumental 
resolution of $R$$\sim$3,200 (except for the top SpeX spectrum which is of $R$$\sim$2500),
continuum normalized, and shifted 
vertically for detailed comparisons of line depths. The left-hand short horizontal 
dashes mark the zero-flux level for each spectrum. The spectrum of $\rho$ Cas in 
outburst best fits the early K-type supergiant 63 Cyg, although with noticeable 
exceptions (see text).
The two outburst spectra are shown to demonstrate that the discussed features
are real and not due to noise. 
}\label{fig3}
\end{figure}

\begin{figure} 
\caption{{\it Crosses}: the visual light curve of $\rho$ Cas from AAVSO binned
in 10$^{d}$ averages. The size of symbols is proportional to the number of averaged datapoints. 
{\it Dashed line}: radial velocity curve from photospheric absorption lines 
of \citet{Lobel2003}. The star velocity is $-$47$\pm$1 $\rm km\,s^{-1}$. 
Arrows show dates of \citet{Sheffer93} CO observations presented with the same color scheme on Fig. 5c.
{\it Upper panel}: our low-resolution $K$ band spectra and one high-resolution 
spectrum from Hinkle et al. (1981) of $\rho$ Cas. Near-IR CO is observed in emission
on the descending branch of the light curve when star begins to expand out of 
minimum radius.
}\label{fig4}
\end{figure}

\begin{figure}  
\caption{a): high-resolution spectra of $\rho$ Cas of the first overtone
$\Delta \nu$2-0 CO band observed in emission in 2004 September ({\em solid line}),
and in absorption in 1979 February ({\em dotted line}; from \citet{Lambert81}).
b): Separate rotational CO transitions of 2004 September are overplotted in the 
velocity scale, together with Na~{\sc i} and Br${\gamma}$ lines observed on the 
same night. 
c): Averaged line profiles of the R14 to R17 CO transitions from 1979 to 1989
observed by \citet{Sheffer93} compared to our observations in 2004.
The spectral resolutions are 9 and 8 $\rm km\,s^{-1}$ per pixel, respectively.
The star velocity of $-$47 $\rm km\,s^{-1}$ is marked with the vertical dashed line.
The spectra have been continuum normalized and shifted in the vertical direction.
The pulsation phases are roughly determined from the light curve in Fig. \ref{fig4},
with $\phi$=0.0 corresponding to maximum brightness, and $\phi$= 0.5 to minimum brightness.
}\label{fig5}
\end{figure}

\begin{figure}
\caption{The spectral energy distribution in the low-resolution CorMASS spectra
can be used to constrain the photospheric parameters of $\rho$ Cas,
together with the temperature-sensitive hydrogen lines and the gravity-sensitive 
Ca~{\sc ii} lines. The graph shows the best-fit model spectra from the NextGen-giant 
grid to the spectrum of 2004 October. The strong calcium triplet lines favor
the lowest gravity value of 0.0, for which models with only T$_{\rm eff} \le 6400$ K are 
available. Near-IR CO is not expected to form in this temperature range confirming 
our calculations using Kurucz model atmospheres (see \S \ref{shock}).
}\label{fig6}
\end{figure}

\begin{figure}
\caption{{\it Upper panel}: the atmospheric gas
temperature structures used to calculate the model spectra. The horizontal axis 
is in the column mass scale $m$. The outer layers of the atmosphere correspond to 
smaller values of log $m$. The marked numbers on the atmospheric temperature profiles 
show the corresponding geometric scale expressed as the height above the photosphere 
(where by definition the optical depth $\tau$ =2/3 and $T_{\rm g}$= $T_{\rm eff}$) 
in R$_{\odot}$ ({\it upper row}), and in percent of the photospheric radius 
(adopting $R_{*}$= 450 R$_{\odot}$ for $\rho$ Cas; {\it lower row}).
{\it Lower panel}: CGS4 high-resolution spectrum of $\rho$ Cas of 
2004 September ({\it solid black line}) plotted against synthetic model spectra
calculated with temperature structures in the upper panel. 
The model spectra are blue-shifted by 7 $\rm km\,s^{-1}$ to match the 
velocity of the observed CO bandhead.
Dot-dashed spectra are calculated using Kurucz atmosphere models,
while the long-, short-dashed and dotted models include a cool gas layer in the upper atmosphere
at log $m$ = $-$3, $-$3.5, and $-$4, respectively. The model adapted with a temperature
minimum at log $m$=$-$3 best fits the observed emission flux around the CO bandhead.
None of the adapted models yield significant NIR CO absorption.
}\label{fig7}
\end{figure}

 \begin{figure}  
\caption{Examples of the main types of atomic spectral lines observed in $\rho$ Cas, 
arranged in progression from pure absorption lines (of hydrogen) to pure emissions lines 
(of sodium). Compare with the line profiles of several $\Delta \nu =$2-0 CO transitions.
The oscillator strengths and excitation energies of the lower and upper energy levels
for a given transition are shown in the lower right corner.
Lines beyond 1 $\mu$m are from the CGS4 echelle spectrum of 2004 September,
while lines shortward of 1 $\mu$m are from the ECHLR spectrum of 2004 October.
The spectra of Arcturus, the Sun, and the telluric spectra are from \citet{atlas}.
The NIR spectra are corrected for telluric absorption; weak telluric lines remain in the optical. 
The spectral resolution of all the NIR spectra and of the optical spectrum
of $\rho$ Cas is $\sim$40,000, whereas for the optical spectra of the Sun and Arcturus 
$R$ $\sim$150,000, which in all cases sufficiently resolves the line profiles for direct comparisons.
}\label{fig8}
 \end{figure}

 \begin{figure}
\caption{High-resolution optical Sofin spectra of $\rho$ Cas during the same pulsation 
cycle as our NIR CO observations. The upper panel shows variations in the
high-excitation Fe~{\sc i} 5572.84 \AA\, absorption line which probes the variable 
photospheric radial velocity and changes of $T_{\rm eff}$. The violet line 
wing forms in the upper atmosphere and probes opacity changes in the supersonic wind
of the hypergiant. The core of the neutral iron line remains single throughout the entire 
pulsation cycle, and increases EW when $T_{\rm eff}$ decreases. The lower panel 
shows three examples of lines that develop emission cores around maximum brightness
(type III lines in our terminology). The flux of the emission lines is clearly 
intrinsically variable, independent of the changes in the underlying 
photospheric absorption spectrum.
}\label{fig9}
\end{figure}

\begin{figure}
\epsscale{0.9}
\plotone{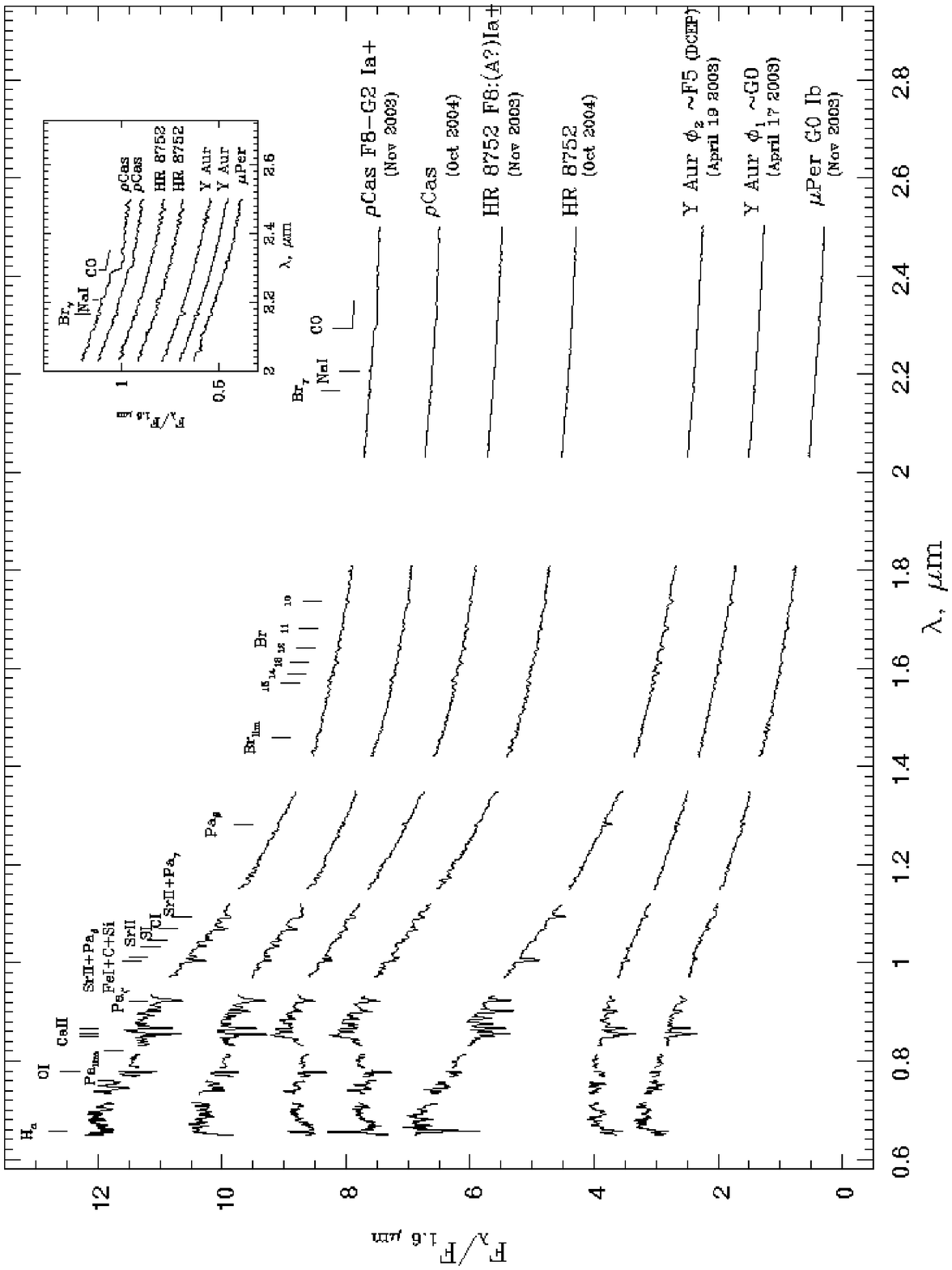}
Fig. 1
\end{figure}

\epsscale{0.9}
\begin{figure}  
\plotone{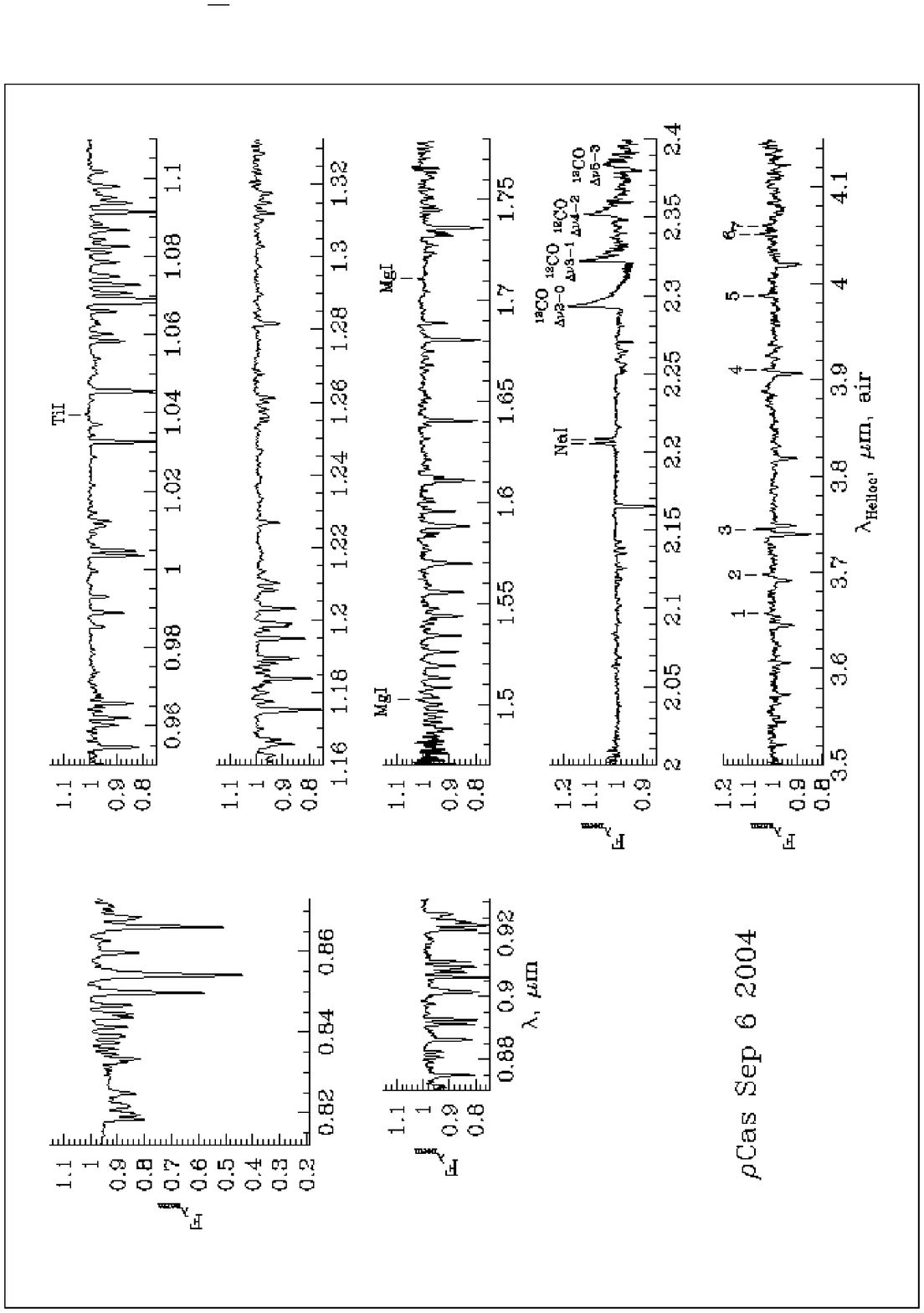}  
Fig. 2
\end{figure}

\epsscale{0.9}
\begin{figure} 
\plotone{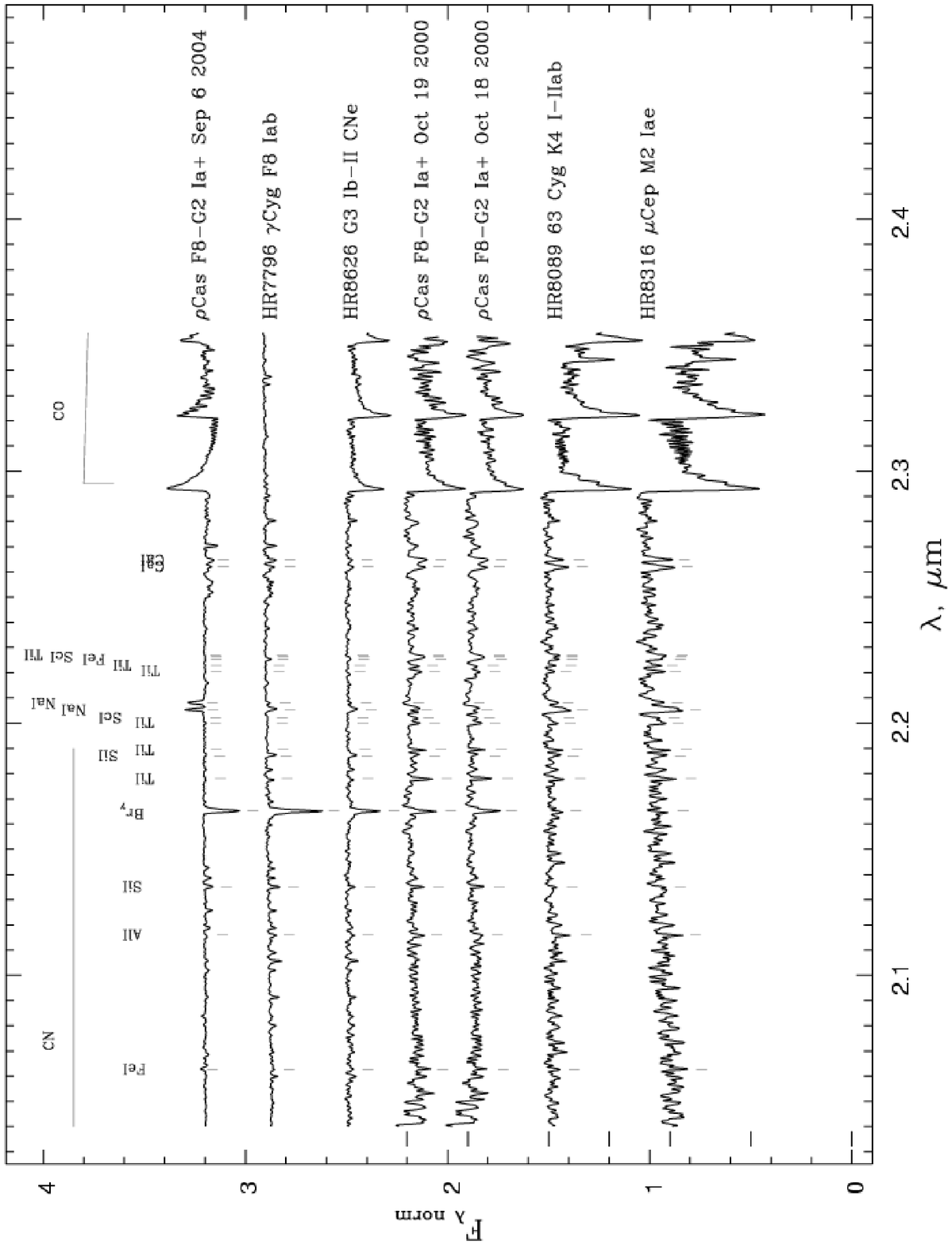} 
Fig. 3
\end{figure}

\begin{figure}
\epsscale{0.9} 
\plotone{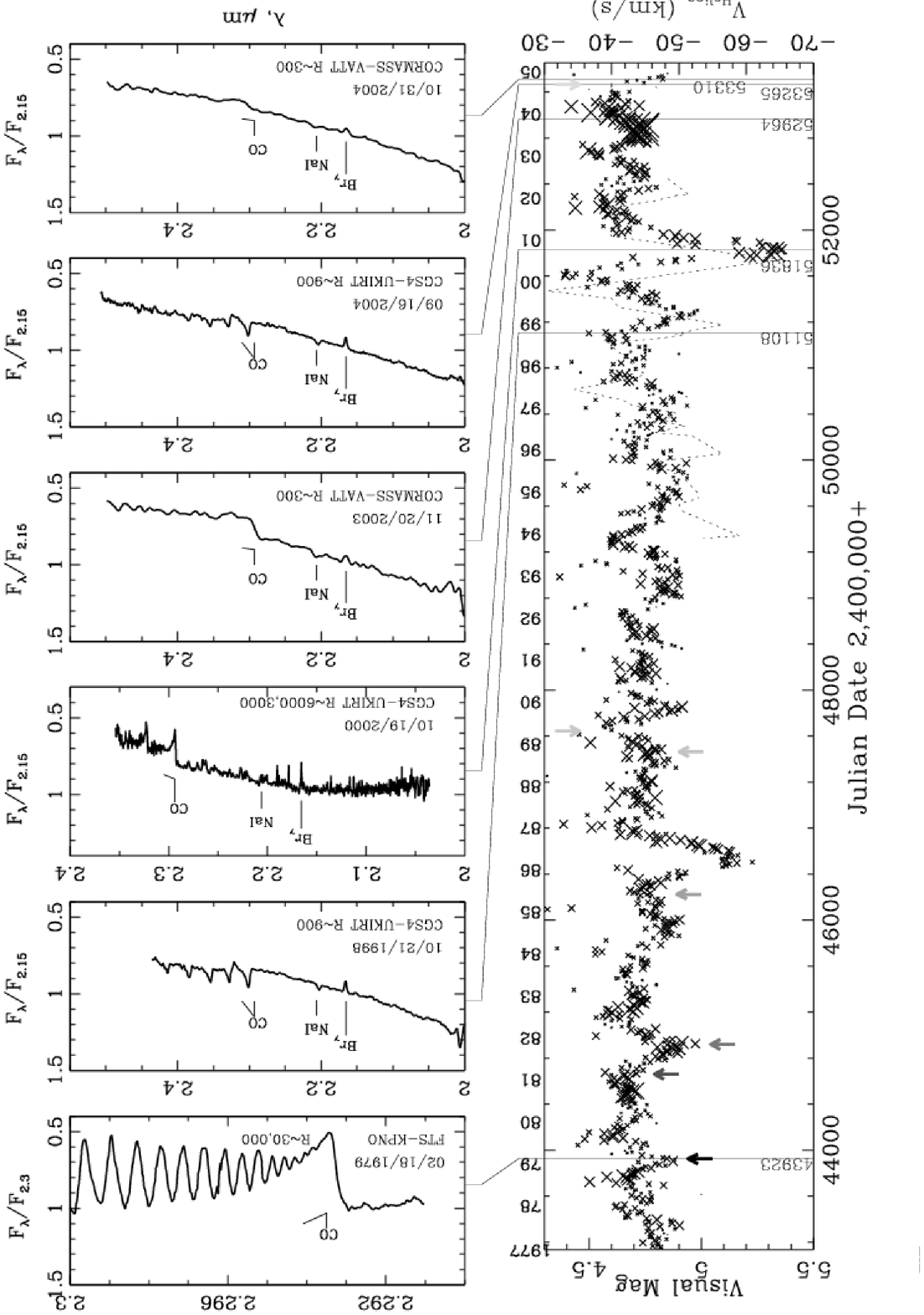} 
Fig. 4
\end{figure}

\begin{figure}  
\epsscale{0.9}
\plotone{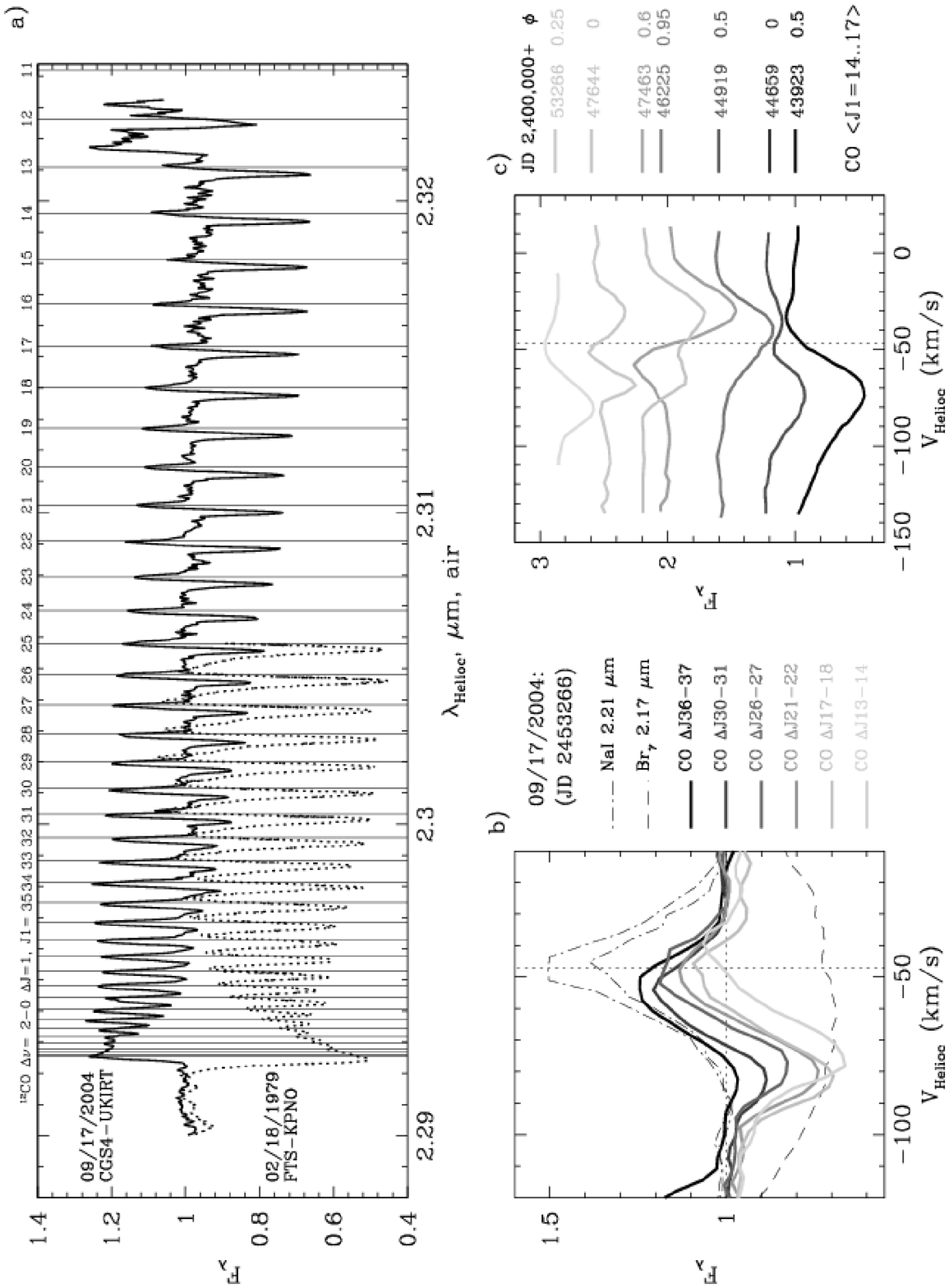} 
Fig. 5
\end{figure}

\begin{figure}
\epsscale{0.9}
\plotone{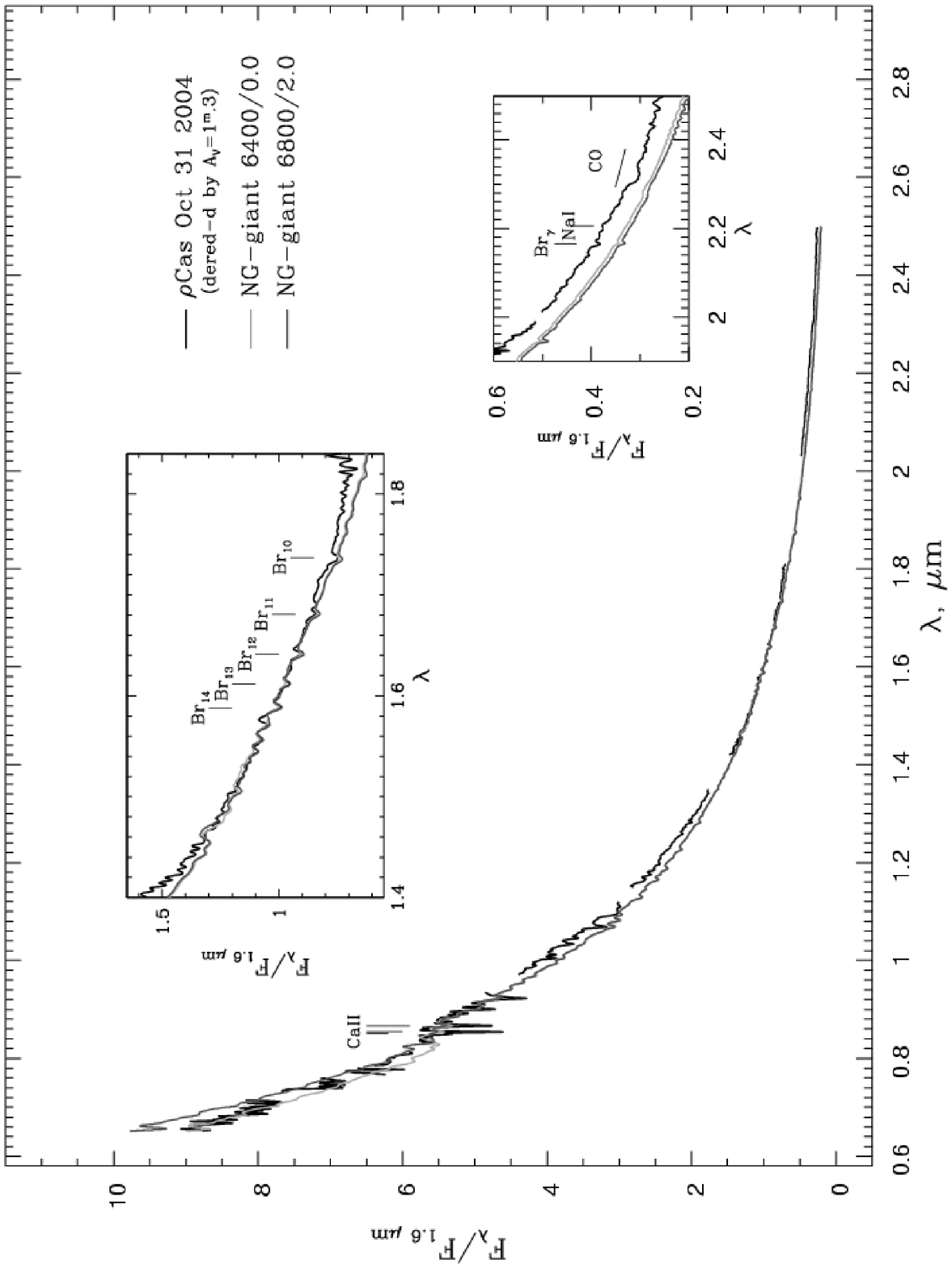}   
Fig. 6
\end{figure}

\begin{figure}
\epsscale{0.9}
\plotone{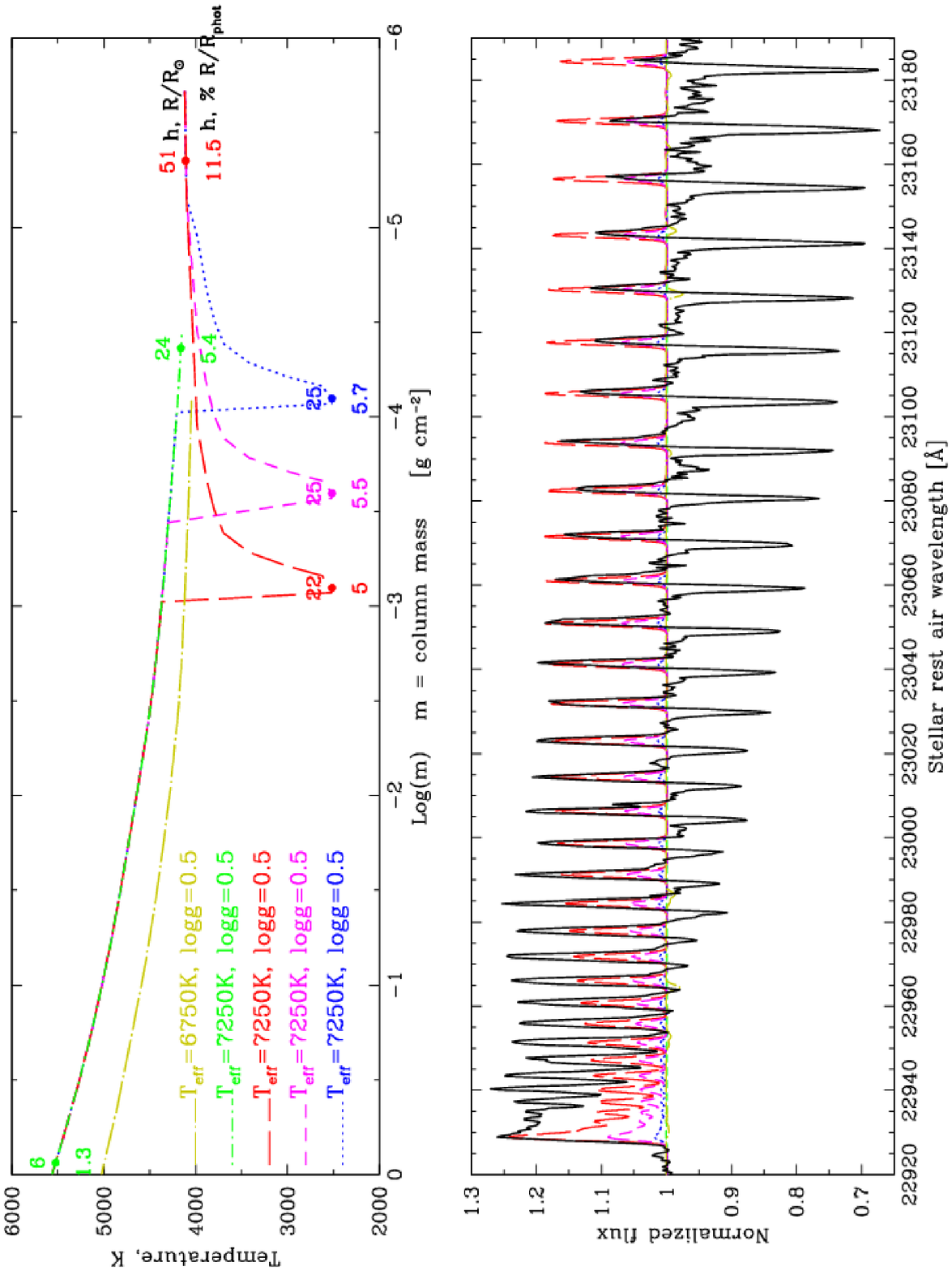}   
Fig. 7
\end{figure}

 \begin{figure}  
\epsscale{0.9}
\plotone{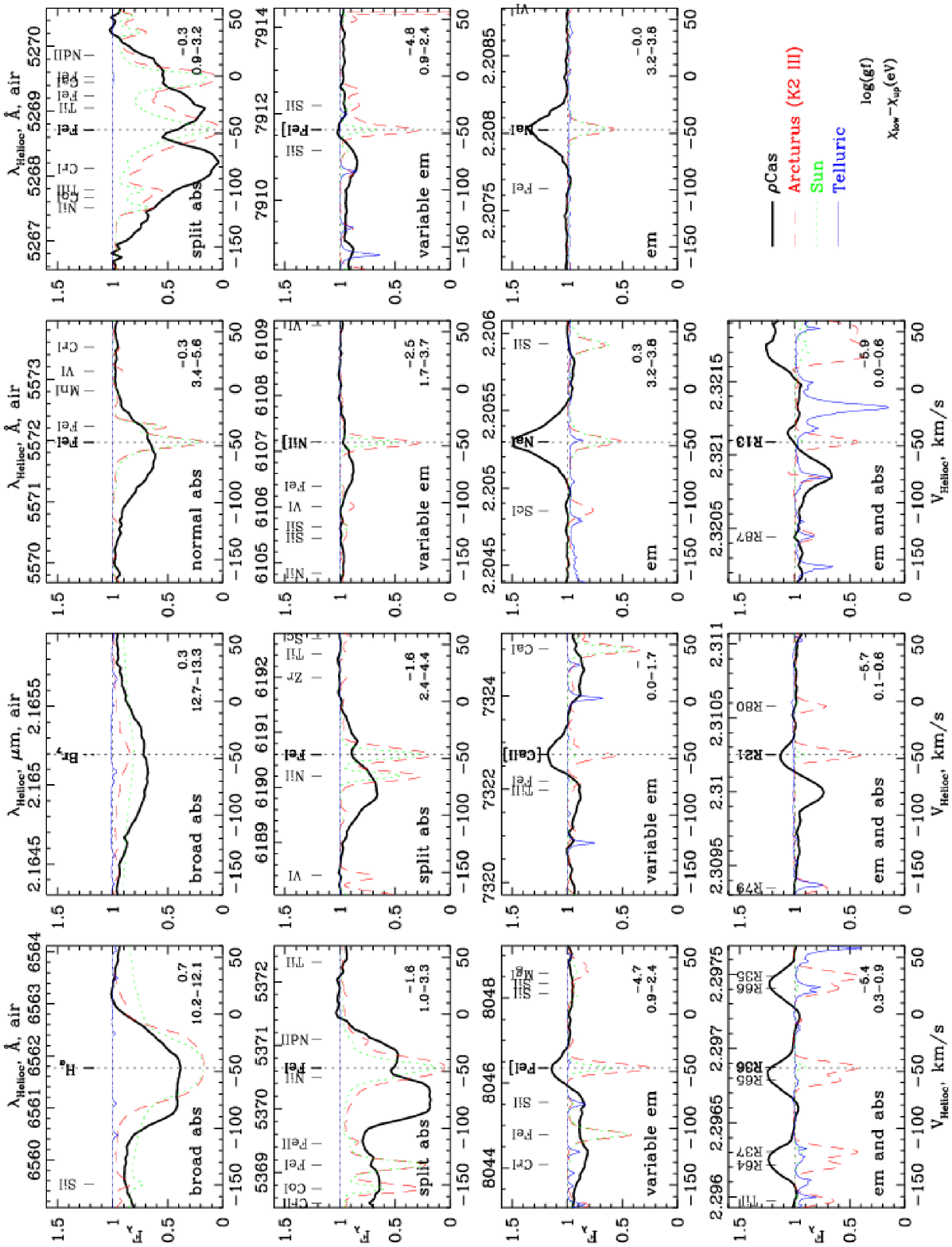} 
Fig. 8
 \end{figure}

 \begin{figure}
\epsscale{1.0}
\plotone{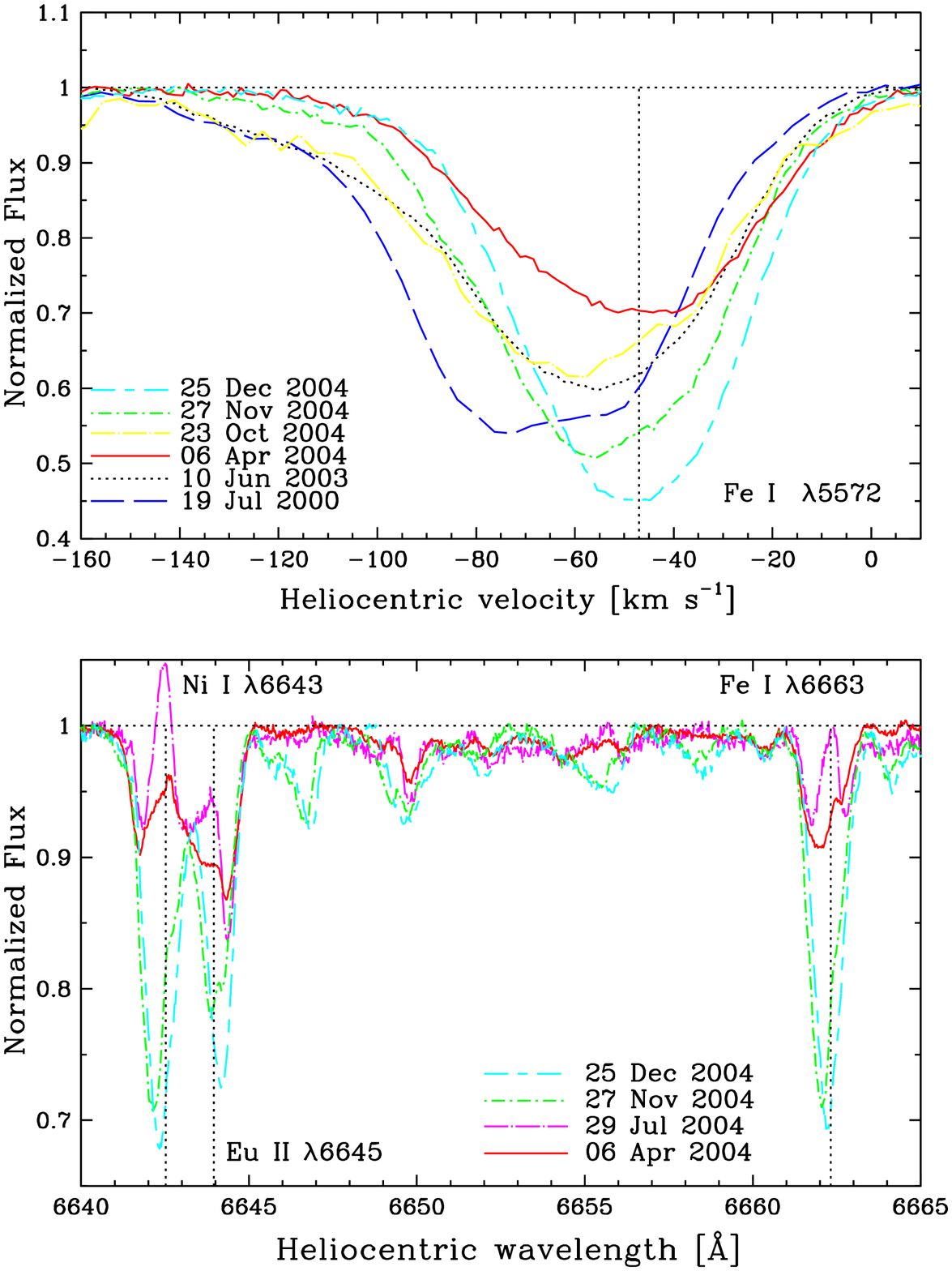}   
Fig. 9
\end{figure}

\clearpage  

\appendix

\section{Spectral Line Identification Lists}\label{lines}

Table \ref{table3} provides wavelengths of spectral lines marked in the 
figures of this paper. Sources of wavelengths we used are:
\begin{itemize}
\item Atomic Line List v. 2.04, compiled by P. van Hoof\\
\texttt{www.pa.uky.edu/$\sim$peter/atomic}
\item 1995 Atomic Line Data by R. L. Kurucz and B. Bell,
Kurucz CD-ROM No. 23. Cambridge, Mass.: Smithsonian Astrophysical Observatory\\
\texttt{cfa-www.harvard.edu/amdata/ampdata/kurucz23/sekur.html}
\item CO Line List by R. L. Kurucz\\
\texttt{kurucz.harvard.edu/LINELISTS/linesmol/coxx.asc}
\end{itemize}

\clearpage  
\begin{deluxetable}{lcr}
\tablecolumns{3} 
\tablewidth{0pt}
\tabletypesize{\small}
\tablecaption{List of observed spectral lines\label{table3}.}
\tablehead{ 
\colhead{$\lambda$ (\AA) \tablenotemark{a}} & \colhead{Element} & \colhead{Figure No.}
}
\startdata

5572.842 & Fe~{\sc i} & 8, 9\\
5269.537 & Fe~{\sc i} & 8\\
5371.489 & Fe~{\sc i} & 8 \\
6191.557 & Fe~{\sc i} & 8 \\
6108.116  & Ni~{\sc i}] &  8 \\
6562.80 & H${\alpha}$ & 1, 8  \\
6643.630 & Ni~{\sc i}]   & 9\\
6645.064 & Eu~{\sc ii}   & 9\\
6663.441 & Fe~{\sc i} & 9\\
7323.89 & [Ca~{\sc ii}] & 8 \\
7771.944, 7774.166, 7775.387 & O~{\sc i} & 1 \\
7912.866 & Fe~{\sc i}] & 8 \\
8047.617 & Fe~{\sc i}] & 8 \\
8203.6 & Pa$_{\rm lim}$ & 1 \\
8498.03, 8542.09, 8662.14 & Ca~{\sc ii} & 1 \\
9229.014  & Pa${\zeta}$ & 1 \\
10036.653 & Sr~{\sc ii} & 1 \\
10049.373  & Pa${\delta}$ & 1\\
10114.015 & Fe~{\sc i} & 1 \\
10123.86 & C~{\sc i} & 1 \\
10124.940 &  Si~{\sc i} & 1 \\
10327.311 & Sr II & 1 \\
10396.76 & Ti~{\sc i} & 2\\
10455.449, 10456.757, 10459.406 & S~{\sc i} & 1 \\
10683.08, 10685.34, 10691.24 & C~{\sc i} & 1 \\
10914.887 & Sr~{\sc ii} & 1 \\
10938.095 & Pa${\gamma}$ & 1  \\
12818.08 & Pa${\beta}$ & 1 \\
14584.2 & Br$_{\rm lim}$ &  1 \\
15024.993 & Mg~{\sc i} & 2\\
15700.662 & Br$_{15}$ & 1 \\
15880.541   & Br$_{14}$ & 1, 6 \\
16109.313   & Br$_{13}$ & 1, 6\\
16407.192    & Br$_{12}$ & 1, 6\\
16806.520 & Br$_{11}$ & 1, 6 \\
17108.663 & Mg~{\sc i} & 2 \\
17362.108     & Br$_{10}$ & 1, 6 \\
20629.757 & Fe~{\sc i} & 3\\
21163.76 & Al~{\sc i} &  3\\
21354.226 & Si~{\sc i} & 3 \\
21655.29    & Br${\gamma}$ & 1, 3, 4, 5, 6, 8\\
21782.7 & Ti~{\sc i} & 3\\
21874.15 & Si~{\sc i} & 3\\
21897.3 & Ti~{\sc i} & 3\\
22004.3 & Ti~{\sc i} & 3\\
22024.3 & Sc~{\sc i}] & 3\\
22056.4 & Na~{\sc i} & 1, 2, 3, 4, 5, 6, 8 \\
22083.7 & Na~{\sc i} & 1, 2, 3, 4, 5, 6, 8 \\
22210.6 & Ti~{\sc i} & 3\\
22232.9  & Ti~{\sc i} & 3\\
22257.108 & Fe~{\sc i} & 3\\
22266.7  & Sc~{\sc i} & 3\\
22274.0  & Ti~{\sc i} & 3\\
22624.97, 22626.73  & Ca~{\sc i} & 3\\
22651.18, 22653.58, 22655.35  & Ca~{\sc i} & 3 \\
22928.976 & $^{12}$CO $\Delta \nu$ 2-0 bandhead &  1, 2, 3, 4, 5, 6, 7\\
22972.326 &$\Delta$J 36-37 & 5, 8\\
23106.100 &$\Delta$J 21-22 & 5, 8\\
23214.513 &$\Delta$J 13-14 & 5, 8\\
23220.222 &$^{12}$CO $\Delta \nu$ 3-1 bandhead &  2\\
23518.163 &$^{12}$CO $\Delta \nu$ 4-2 bandhead &  2\\
23822.970 &$^{12}$CO $\Delta \nu$ 5-3 bandhead &  2\\ 

\enddata
\tablenotetext{a}{Rest wavelength in air.}
\end{deluxetable}


\clearpage  

\begin{thebibliography}{}

\bibitem[Alvarez et al.(2001)]{Alvarez01} Alvarez, R., Jorissen, A., Plez, B., Gillet, D., Fokin, A., \& Dedecker, M. 2001, \aap, 379, 305 
\bibitem[Arellano Ferro(1985)]{Arrelano1985} Arellano Ferro, A. 1985, \mnras, 216, 571
\bibitem[Ayres(2002)]{Ayres02} Ayres, T. R. 2002, \apj, 575, 1104
\bibitem[Banerjee \& Ashok(2002)]{Banerjee02}  Banerjee, D. P. K. \& Ashok, N. M. 2002, \aap, 395, 161
\bibitem[Bell, Edvardsson, \& Gustafsson(1985)]{bell85} Bell, R. A., Edvardsson, B., Gustafsson, B. 1985, \mnras, 212, 497
\bibitem[Bessel, Scholz, \& Wood(1996)]{Bessel96} Bessel, M. S., Scholz, M., \& Wood, P. R.  1996, \aap, 307, 481
\bibitem[Bidelman \& McKellar(1957)]{Bidelman55} Bidelman, W. P. \& McKellar, A.  1957, \pasp, 69, 31
\bibitem[Bieging, Rieke, \& Rieke(2002)]{Bieging02} Bieging, J. H., Rieke, M. J., \& Rieke, G. H. 2002,
        \aap, 384, 965
\bibitem[Biscaya et al.(1997)]{Biscaya97} Biscaya, A. M., Rieke, G. H., Narayanan, G., Luhman, K. L., \& Young, E. T. 1997,
        \apj, 491, 359
\bibitem[Chadid \& Gillet(1996)]{Chadid96} Chadid, M. \& Gillet, D. 1996, \aap, 308, 481
\bibitem[Clayton et al.(1994)]{Clayton94} Clayton, G. C., Lawson, W. A., Cottrell, P. L., Whitney, B. A., Stanford, S. A., \& de Ruyter, F.  1994, \apj, 432, 785 
\bibitem[Clayton(1996)]{Clayton96} Clayton, G. C.  1996, \pasp, 108, 225
\bibitem[Clayton, Geballe, \& Bianchi(2003)]{Clayton03} Clayton, G. C., Geballe, T. R., \& Bianchi, L.  2003, \apj, 595, 412
\bibitem[Cottrell \& Lambert(1982)]{Cottrell82} Cottrell, P. L. \& Lambert, D. L.  1982, Observatory, 102, 149
\bibitem[Cushing, Vacca \& Rayner(2004)]{Cushing04} Cushing, M. C., Vacca, W. D., \& Rayner, J. T.  2004, \pasp, 116, 362
\bibitem[de Jager, Lobel \& Israelian(1997)]{deJager97} de Jager, C., Lobel, A., \& Israelian, G. 1997,
        \aap, 325, 714
\bibitem[de Jager(1998)]{deJager98}  de Jager, C.  1998, \aapr, 8, 145 
\bibitem[Evans et al.(1996)]{Evans96} Evans, A., Geballe, T. R., Rawlings, J. M. C., \& Scott, A. D. 1996,
        \mnras, 282, 1049
\bibitem[Ferland et al.(1979)]{Ferland79} Ferland, G. J., Lambert, D. L., Netzer, H., Hall, D. N. B., \& Ridgway, S. T. 1979, \apj, 227, 489
\bibitem[Fix \& Cobb(1987)]{Fix87} Fix, J. D. \& Cobb, M. L.  1987, \apj, 312, 290 
\bibitem[Gesicki(1992)]{Gesicki92} Gesicki, K.  1992, \aap, 254, 280
\bibitem[Gillet et al.(1989)]{Gillet89} Gillet, D.,  Duquennoy, A., Bouchet, P., \& Gouiffes, C.  1989, \aap, 215, 316 
\bibitem[Gonzalez et al.(1998)]{Gonzalez98} Gonzalez, G., Lambert, D. L., Wallerstein, G., Rao, N. K., Smith, V. V.,
        McCarthy, J. K.  1998, \apjs, 114, 133
\bibitem[Gottlieb \& Liller(1978)]{Gottlieb78} Gottlieb, E. W. \& Liller, W.  1978, \apj, 225, 488
\bibitem[Harmer, Lawson \& Stickland(1978)]{Harmer78} Harmer, D. L., Lawson, P. A., \& Stickland, D. J. 1978, Obs, 98, 250
\bibitem[Hinkle, Hall, \& Ridgway(1982)]{Hinkle82}  Hinkle, K. H., Hall, D. N. B., \& Ridgway, S. T.  1982, \apj, 252, 697
\bibitem[Hinkle, Joyce, \& Smith(1995)]{Hinkle95} Hinkle, K. H., Joyce, R. R., \& Smith, V.  1995, \aj, 109, 808
\bibitem[Hinkle et al.(2003)]{atlas} Hinkle, K., Wallace, L., Livingston, W., Ayres, T., Harmer, D., \& Valenti, J.  2003, 
        "High Resolution Infrared, Visible and Ultraviolet Spectral Atlases of the Sun and Arcturus" in 12th Cambridge
        Workshop on Cool Stars, Stellar Systems and the Sun, University of Colorado, eds. A. Brown, G .M. Harper, and T. R. Ayres;
        available in the digital form at http://www.archive.noao.edu and ftp://ftp.noao.edu/catalogs/hiresK
\bibitem[Hauschildt et al.(1999)]{Hauschildt99} Hauschildt, P. H., Allard, F., Ferguson, J., Baron, E., \& Alexander, D. R.  1999, \apj, 525, 871
\bibitem[Hrivnak, Kwok \& Geballe(1994)]{Hrivnak94} Hrivnak, B. J., Kwok, S., \& Geballe, T. R.  1994, \apj, 420, 783
\bibitem[Humphreys(1978)]{Humphreys78} Humphreys, R. M.  1978, \apjs, 38, 309
\bibitem[Humphreys, Davidson \& Smith(2002)]{Humphreys02} Humphreys, R. M., Davidson, K., \& Smith, N.  2002, \aj, 124, 1026
\bibitem[Ilyin(2000)]{Ilyin2000} Ilyin, I.  2000, Ph.D. thesis (Univ. of Oulu, Finland)
\bibitem[Israelian, Lobel \& Schmidt(1999)]{Israelian99} Israelian, G., Lobel, A., \& Schmidt, M. R. 1999, 523, L145
\bibitem[Jura \& Kleinmann(1990)]{Jura90} Jura, M. \& Kleinmann, S. G. 1990, \apj, 351, 583
\bibitem[Kameswara Rao \& Lambert(1997)]{Rao97} Kameswara Rao, N. \& Lambert, D. L.  1997, \mnras, 284, 489
\bibitem[Kameswara Rao et al.(1999)]{Rao99} Kameswara Rao, N., Lambert, D. L., Adams, M. T., Doss, D. R., Gonzalez, G.,
                Hatzes, A. P., James, C. R., Johns-Krull, C. M., Luck, R. E., Pandey, G., Reinsch, K., Tomkin, J., \&  Woolf, V. M.  1999, \mnras, 310, 717
\bibitem[Kleinmann \& Hall(1986)]{Kleinmann86} Kleinmann, S. G. \& Hall, D. N. B. 1986, \apjs, 62, 501 
\bibitem[Kovtyukh et al.(2003)]{Kovtyukh03} Kovtyukh, V. V., Andrievsky, S. M., Luck, R. E., \& Gorlova, N. I.  2003, \aap, 401, 661
\bibitem[Kovtyukh \& Gorlova(2000)]{Kovtyukh2000} Kovtyukh, V. V. \& Gorlova, N. I. 2000, \aap, 358, 587
\bibitem[Kraus et al.(2000)]{Kraus00} Kraus, M., Kraegel, E., Thum, C., \& Geballe, T. R. 2000, \aap, 362, 158
\bibitem[Lambert \& Luck(1978)]{Lambert78} Lambert, D. L. \& Luck, R. E.  1978, \mnras, 184, 405
\bibitem[Lambert, Hinke \& Hall(1981)]{Lambert81} Lambert, D. L., Hinkle, K. H., \& Hall, D. N. B. 1981, \apj, 248, 638
\bibitem[Lejeune \& Schaerer(2001)]{Lejeune2001} Lejeune, T. \& Schaerer, D. 2001, \aap, 366, 538
\bibitem[Lobel et al.(1994)]{Lobel94} Lobel, A., de Jager, C., Nieuwenhuijzen, H., Smolinski, J., \& Gesicki, K. 1994, \aap, 291, 226
\bibitem[Lobel(1997)]{Lobel1997} Lobel, A. 1997, Pulsation and Atmospherical Instability of Luminous F- and G-type Stars; The Yellow Hypergiant $\rho$ Cassiopeiae, ISBN 90-423-0014-0, Shaker Publ., Maastricht, The Netherlands  
\bibitem[Lobel et al.(1998)]{Lobel98} Lobel, A., Israelian, G., de Jager, C., Musaev, F., Parker, J. Wm., \& Mavrogiorgou, A. 1998, \aap, 330, 659
\bibitem[Lobel(2001)]{Lobel2001} Lobel, A. 2001, \apj, 558, 780 
\bibitem[Lobel et al.(2003)]{Lobel2003} Lobel, A.,  Dupree, A. K., Stefanik, R. P., Torres, G., Israelian, G., Morrison, N., de Jager, C., Nieuwenhuijzen, H., Ilyin, I., \& Musaev, F. 2003, \apj, 583, 923
\bibitem[Lobel(2004)]{Lobel2004} Lobel, A., 2004, Mercury,  ASP, Vol. 33, 1, 13
\bibitem[Lobel et al.(2005)]{Lobel2005} Lobel, A., Aufdenberg, J., Ilyin, I., \& Rosenbush A. E. 2005, ``Mass-loss and Recent Spectral Changes in the Yellow Hypergiant $\rho$ Cas'' in 13th Cambridge Workshop on Cool Stars, Stellar Systems and the Sun, ESA SP, eds. F. Favata et al., in press (astro-ph/0504524)
\bibitem[Lynch et al.(2004)]{Lynch2004} Lynch, D. K., Rudy, R. J., Russell, R. W., Mazuk, S., Venturini, C. C., Dimpfl, W.,
        Bernstein, L. S., Sitko, M. L., Fajardo-Acosta, S., Tokunaga, A., Knacke, R., Puetter, R. C., \& Perry, R. B.  2004, \apj, 607, 460
\bibitem[Matsuura et al.(2002)]{Matsuura02} Matsuura, M., Yamamura, I., Zijlstra, A. A., \& Bedding, T. R.  2002, \aap, 387, 1022
\bibitem[McGovern et al.(2004)]{McGovern04} McGovern, M. R., Kirkpatrick, J. D., McLean, I. S., Burgasser, A. J., Prato, L., \& Lowrance, Patrick J. 2004, \apj, 600, 1020
\bibitem[McGregor, Hyland, \& Hillier(1988a)]{McGregor1988a} McGregor, P. J., Hyland, A. R., \& Hillier, D. J.  1988a, \apj, 324, 1071
\bibitem[McGregor, Hyland, \& Hillier(1988b)]{McGregor1988}  McGregor, P. J., Hyland, A. R., \& Hillier, D. J.  1988b, \apj, 334, 639
\bibitem[Mountain et al.(1990)]{Mountain90} Mountain C. M., Robertson D. J., Lee T. J., Wade R., 1990, 
        in Crawford D. L., ed., Proc. SPIE Vol. 1235, Instruments in Astronomy VII. SPIE, Bellingham, p. 25
\bibitem[Mozurkewich et al.(1987)]{Mozurkewich87} Mozurkewich, D., Gehrz, R. D., Hinkle, K. H., Lambert, D. L.  1987, \apj, 314, 242
\bibitem[Nieuwenhuijzen \& de Jager(1995)]{Nieuwenhuijzen1995} Nieuwenhuijzen, H. \& de Jager, C. 1995, \aap, 302, 811
\bibitem[Nieuwenhuijzen \& de Jager(2000)]{Nieuwenhuijzen00} Nieuwenhuijzen, H. \& de Jager, C. 2000, \aap, 353, 163
\bibitem[Nowotny et al.(2005a)]{Nowotny05a} Nowotny, W., Aringer, B.,  H\"{o}fner, S., Gautschy-Loidl, \& Windsteig, W.  2005a,
        \aap, 437, 273
\bibitem[Nowotny et al.(2005b)]{Nowotny05b} Nowotny, W., Lebzelter, T., Hron, J., H\"{o}fner 2005b, \aap, 437, 285
\bibitem[Oudmaijer et al.(1994)]{Oudmaijer94} Oudmaijer, R. D., Geballe, T. R., Waters, L. B. F. M., \& Sahu, K. C.  1994, \aap, 281, L33
\bibitem[Oudmaijer et al.(1995)]{Oudmaijer95} Oudmaijer, R. D., Waters, L. B. F. M., 
       van der Veen, W. E. C. J., \& Geballe, T. R.  1995, \aap, 299, 69
\bibitem[Pandey et al.(2004)]{Pandey04} Pandey, G., Lambert, D. L., Rao, N. K., Gustafsson, B., Ryde, N., Yong, D.  2004, \mnras, 353, 143 
\bibitem[Percy \& Zsoldos(1992)]{Percy92} Percy, J. R. \& Zsoldos, E.  1992, \aap, 263, 123
\bibitem[Perrin et al.(2004)]{Perrin04} Perrin, G., Ridgway, S. T., Coud\'e du Foresto, V.,
        Mennesson, B., Traub, W. A., \& Lacasse, M. G. 2004, \aap, 418, 675
\bibitem[Rayner et al.(2003)]{Rayner03} Rayner, J. T., Toomey, D. W., Onaka, P. M., Denault,  A. J., Stahlberger,  W. E., Vacca,  W. D., Cushing, M. C.,
       \& Wang, S.  2003, \pasp, 115, 362
\bibitem[Richter et al.(2003)]{Richter03} Richter, He., Wood, P. R., Woitke, P., Bolick, U., \& Seldmayr, E.  2003, \aap, 400, 319
\bibitem[Ridgway et al.(1984)]{Ridgway84} Ridgway, S. T., Carbon, D. F., Hall, D. N. B., \& Jewell, J. 1984, \apjs, 54, 177 
\bibitem[Rudy et al.(2003)]{Rudy03} Rudy, R. J., Dimpfl, W. L., Lynch, D. K., Mazuk, S., Venturini, C. C.,
       Wilson, J. C., Puetter, R. C., \& Perry, R. B. 2003, \apj, 596, 1229
\bibitem[Rushton et al.(2005)]{Rushton05} Rushton, M. T., Geballe, T. R., Filippenko, A. V., Chornock, R., Li, W., Leonard, D. C., Foley, R. J.,
               Evans, A., Smalley, B., van Loon, J. Th., \& Eyres, S. P. S.  2005, \mnras, in press 
\bibitem[Samus et al.(2004)]{Samus04} Samus, N. N. et al.  2004, Combined General Catalogue of Variable Stars (Institute of Astronomy, Moscow), available
            as on-line VizieR catalog II/250 (GCVS)
\bibitem[Sanford(1952)]{Sanford52} Sanford, R. F.  1952, \apj, 116, 331
\bibitem[Sargent(1961)]{Sargent61} Sargent, W. L. W.  1961, \apj, 134, 142
\bibitem[Schuster, Humphreys \& Marengo(2006)]{Schuster05}  Schuster, M. T., Humphreys, R. M., \& Marengo, M.  2006,
\aj, 131, 603
\bibitem[Sheffer \& Lambert(1986)]{Sheffer86} Sheffer, Y. \& Lambert, D. L.  1986, \pasp, 98, 914
\bibitem[Sheffer \& Lambert(1987)]{Sheffer87} Sheffer, Y. \& Lambert, D. L.  1987, \pasp, 99, 1277
\bibitem[Sheffer(1993)]{Sheffer93} Sheffer, Y. 1993, Behavioral Study of Yellow Supergiants, Ph.D. thesis (The Univ. of Texas, Austin)
\bibitem[Skiff(2005)]{Skiff05} Skiff, B. A.  2005, Catalogue of Stellar Spectral Classifications (Lowell Observatory), available
	    as on-line VizieR catalog III/233B
\bibitem[Skuljan \& Cottrell(2002)]{Skuljan02} Skuljan, Lj., \& Cottrell, P. L. 2002, \mnras, 335, 1133 
\bibitem[Takeda \& Takada-Hidai(1994)]{Takeda94} Takeda, Y. \& Takada-Hidai, M.  1994, PASJ, 46, 395  
\bibitem[Tenenbaum et al.(2005)]{Tenenbaum05} Tenenbaum, E. D.,  Clayton, G. C., Asplund, M., Engelbracht, C. W.,
Gordon, K. D.,  Hanson, M. M.,  Rudy, R. J., Lynch, D. K., Mazuk, S.,  Venturini, C. C., \& Puetter, R. C.  2005, \aj, 130, 256 
\bibitem[Thompson \& Boroson(1977)]{Thompson77} Thompson, R. I. \& Boroson, T. A.  1977, \apj, 216, 75
\bibitem[Tsuji(1988)]{Tsuji88} Tsuji, T.  1988, \aap, 197, 185
\bibitem[Tsuji(2001)]{Tsuji01} Tsuji, T.  2001, \aap, 376, L1 
\bibitem[Vacca, Cushing \& Rayner(2003)]{Vacca03} Vacca, W. D., Cushing, M. C., \& Rayner  2003, \pasp, 115, 389
\bibitem[Wallace \& Hinkle(1996)]{Wallace96} Wallace, L. \& Hinkle, K. 1996, \apjs, 107, 312
\bibitem[Wallace \& Hinkle(1997)]{Wallace97} Wallace, L. \& Hinkle, K. 1997, \apjs, 111, 445, 
       spectra available in the digital form at ftp://ftp.noao.edu/catalogs/medresIR/K\_band/
\bibitem[Wallace \& Hinkle(2002)]{Wallace02} Wallace, L. \& Hinkle, K. 2002, \aj, 124, 3393
\bibitem[Whitney(1962)]{Whitney62} Whitney, C. A. 1962, \apj, 136, 674
\bibitem[Wilson et al.(2001)]{Wilson01} Wilson, J. C., Skrutskie, M. F., Colonno, M. R., Enos, A. T., Smith, J. D.,
       Henderson, C. P., Gizis, J. E., Monet, D. G., \& Houck, J. R.  2001, \pasp, 113, 227
\bibitem[Winters et al.(2000)]{Winters00} Winters, J. M., Keady, J. J., Gauger, A., \& Sada, P. V.  2000, \aap, 359, 651
\bibitem[Woitke, Goeres, \& Sedlmayr(1996)]{Woitke96} Woitke, P., Goeres, A., \& Sedlmayr, E.  1996, \aap, 313, 217
\bibitem[Zsoldos \& Percy(1991)]{Zsoldos91} Zsoldos, E. \& Percy, J. R. 1991, \aap, 246, 441

\end{thebibliography}
\end{document}